\documentclass[aps,prd,twoside,twocolumn,nofootinbib,10pt,showpacs,floatfix]{revtex4-1}
\usepackage{amsmath,amssymb}
\usepackage{graphicx,bm}
\usepackage{slashed}
\usepackage{epstopdf}
\usepackage{ulem} 
\usepackage[usenames]{color}
\usepackage{float}
\usepackage{hyperref}
\usepackage{subfigure}
\usepackage{subfigure}
\usepackage{rotating}
\usepackage{color}
\usepackage{multirow}
\usepackage{dcolumn}
\usepackage{overpic}
\usepackage{booktabs}
\usepackage{makecell}
\usepackage{diagbox}

\renewcommand\sout{\bgroup \color{red} \ULdepth=-.5ex \ULset}

\newsavebox{\tablebox}
\begin{document}

\title{Probing the electromagnetic properties of the $\Sigma_c^{(*)}D^{(*)}$-type doubly charmed molecular pentaquarks}
\author{Hong-Yan Zhou$^{1,2}$}
\email{zhouhy20@lzu.edu.cn}
\author{Fu-Lai Wang$^{1,2}$}
\email{wangfl2016@lzu.edu.cn}
\author{Zhan-Wei Liu$^{1,2,3}$}
\email{liuzhanwei@lzu.edu.cn}
\author{Xiang Liu$^{1,2,3}$\footnote{Corresponding author}}
\email{xiangliu@lzu.edu.cn}
\affiliation{$^1$School of Physical Science and Technology, Lanzhou University, Lanzhou 730000, China\\
$^2$Research Center for Hadron and CSR Physics, Lanzhou University and Institute of Modern Physics of CAS, Lanzhou 730000, China\\
$^3$Lanzhou Center for Theoretical Physics, Key Laboratory of Theoretical Physics of Gansu Province, and Frontiers Science Center for Rare Isotopes, Lanzhou University, Lanzhou 730000, China}

\begin{abstract}
 In this work, we discuss the electromagnetic properties of the $S$-wave $\Sigma_c^{(*)}D^{(*)}$-type doubly charmed molecular pentaquarks, which have close relation to their inner structures. Both the $S$-$D$ wave mixing effect and the coupled channel effect are taken into account. We analyze the magnetic moments for the $S$-wave $\Sigma_c^{(*)}D^{(*)}$ molecular states, and extend our theoretical framework to study the transition magnetic moments and the radiative decay widths between the $S$-wave $\Sigma_c^{(*)}D^{(*)}$ molecules. We discuss the relations between the magnetic moments and the transition magnetic moments for the $S$-wave $\Sigma_c^{(*)}D^{(*)}$-type doubly charmed molecular pentaquarks, which can be regarded as an indirect way to measure their magnetic moments experimentally. The present study may inspire experimentalist's interest in measuring the electromagnetic properties of the hadronic molecular states.
\end{abstract}

\maketitle

\section{Introduction}\label{sec1}

The study of the exotic hadronic states is one of the most attractive and active branches of the hadron physics. In the past decades, a large number of exotic hadronic states have been observed by different experimental collaborations benefited from the accumulation of more and more experimental data with high precision \cite{Brambilla:2019esw}. Since the masses of some observed exotic hadronic states are close to the thresholds of the correspond two hadronic states, explaining these exotic hadronic states as the hadronic molecular states was proposed  \cite{Chen:2016qju,Liu:2013waa,Hosaka:2016pey,Liu:2019zoy,Olsen:2017bmm,Guo:2017jvc,Chen:2022asf,Meng:2022ozq}. Especially, the LHCb Collaboration updated experimental result of the $P_c$ states in 2019 \cite{Aaij:2019vzc}, which provides strong evidence to support the existence of the hidden-charm molecular pentaquark states \cite{Li:2014gra, Wu:2010jy, Karliner:2015ina, Wang:2011rga, Yang:2011wz, Wu:2012md, Chen:2015loa}. Although there exists abundant experimental and theoretical investigations in the past 19 years \cite{Brambilla:2019esw,Chen:2016qju,Liu:2013waa,Hosaka:2016pey,Liu:2019zoy,Olsen:2017bmm,Guo:2017jvc,Chen:2022asf,Meng:2022ozq}, the properties of these observed exotic hadronic states are waiting to be explored in future, and the study of the electromagnetic properties may provide the crucial information to understand the inner structures of these observed exotic hadronic states, which can be used to distinguish different assignments to them.

For the hadronic molecular states, their mass spectrums, strong decay properties, and production processes have been studied extensively in the past decades \cite{Chen:2016qju,Liu:2013waa,Hosaka:2016pey,Liu:2019zoy,Olsen:2017bmm,Guo:2017jvc,Chen:2022asf,Meng:2022ozq}. However, our knowledge of their electromagnetic properties is still not enough. As the fundamental properties and the important physical observable quantities of the hadronic molecular state, the magnetic moments and the transition magnetic moments  may play a vital role in mapping out the inner structures of the discussed hadronic molecular states. In addition, the magnetic moment $\mu$ is related to the magnetic form factor $G_M(Q^2)$ of the hadronic states, and can be obtained by the extrapolation of the magnetic form factor to zero momentum transfer $Q$, which can provide the valuable hints to shed light on how the hadronic constituents are distributed and may deepen our understanding of the nonperturbative behavior of the quantum chromodynamics (QCD).

\begin{figure}[!htbp]
\centering
\begin{tabular}{c}
\includegraphics[width=0.47\textwidth]{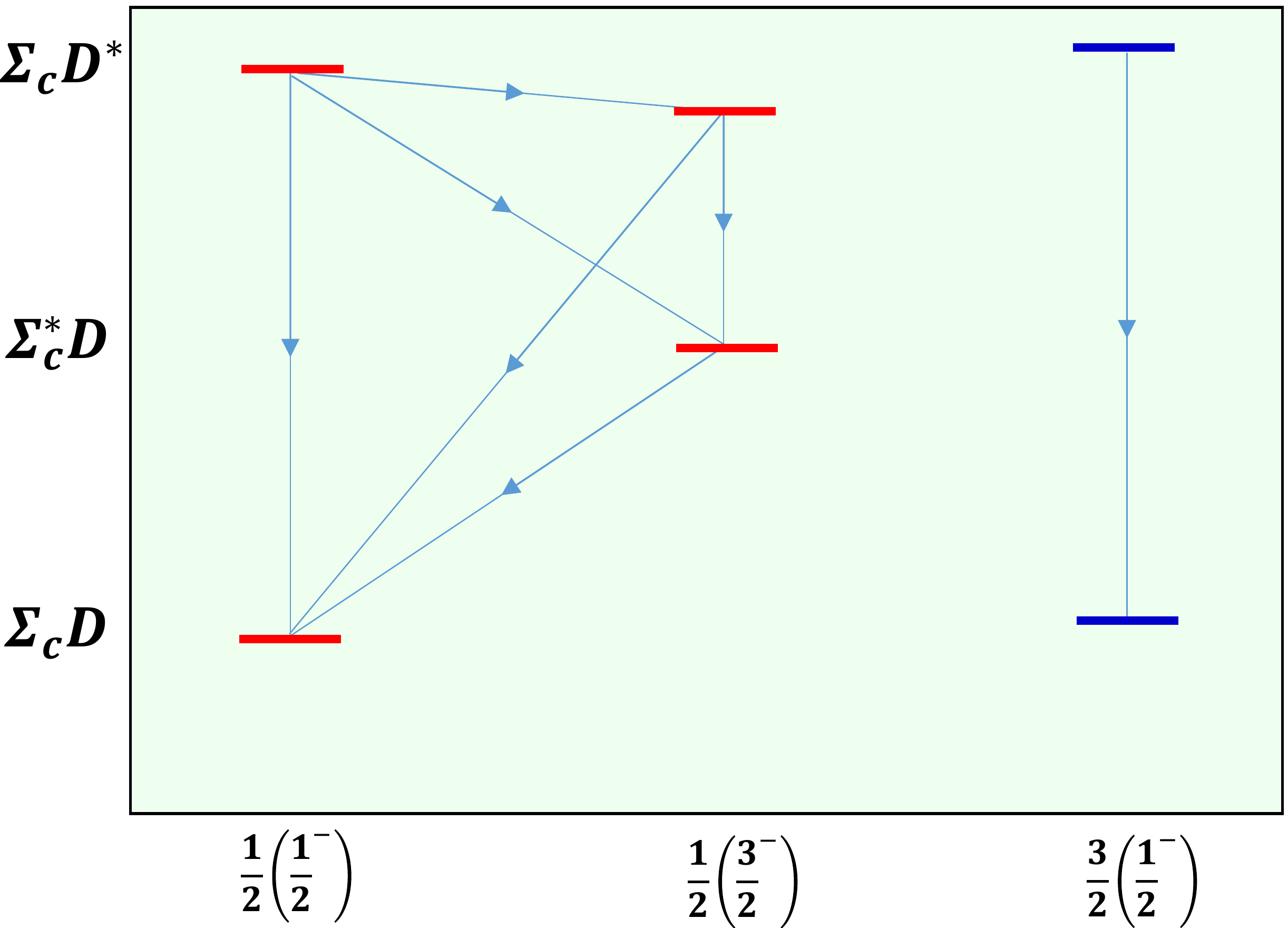}
\end{tabular}
\caption{Summary of the most promising and possible $\Sigma_c^{(*)}D^{(*)}$-type doubly charmed molecular pentaquark candidates \cite{Chen:2021kad}. Here, the lines with arrow represent the possible radiative decay processes of these doubly charmed molecular pentaquark candidates, and the red and blue thick solid lines stand for these most promising and possible doubly charmed molecular pentaquark candidates, respectively.}\label{candidates}
\end{figure}

The observation of the hidden-charm molecular pentaquark states $P_c/P_{cs}$ and the doubly charmed molecular tetraquark state $T_{cc}^+$ \cite{Chen:2016qju,Liu:2013waa,Hosaka:2016pey,Liu:2019zoy,Olsen:2017bmm,Guo:2017jvc,Chen:2022asf,Meng:2022ozq}  naturally makes us conjecture that there should exist the doubly charmed molecular pentaquark states composed of the $S$-wave charmed baryon $\Sigma_c^{(*)}$ and the $S$-wave charmed meson $D^{(*)}$ in the hadron spectroscopy. Recently, the mass spectrum of the $S$-wave $\Sigma_c^{(*)}D^{(*)}$-type doubly charmed molecular pentaquarks was obtained by the one-boson-exchange model, where the $S$-$D$ wave mixing effect and the coupled channel effect were considered \cite{Chen:2021kad}. The $\Sigma_c D$ state with $I(J^P)=1/2(1/2^-)$, the $\Sigma_c^* D$ state with $I(J^P)=1/2(3/2^-)$, and the $\Sigma_c D^*$ states with $I(J^P)=1/2(1/2^-,\,3/2^-)$ can be viewed as the most promising doubly charmed molecular pentaquark candidates, while the $\Sigma_c D$ state with $I(J^P)=3/2(1/2^-)$ and the $\Sigma_c D^*$ state with $I(J^P)=3/2(1/2^-)$ are the possible doubly charmed molecular pentaquark candidates (see Fig. \ref{candidates} for more details).

In order to further reveal the properties of these doubly charmed molecular pentaquark candidates, we attempt to explore their magnetic moments, transition magnetic moments, and radiative decay behaviors in this work. In the past several decades, the magnetic moments for the hadronic states have been studied within various models and approaches \cite{Meng:2022ozq}, where the constituent quark model was 
popularly applied to describe the magnetic moments of the decuplet and octet baryons quantitatively \cite{Schlumpf:1993rm,Ramalho:2009gk}. Considering the success of applying the constituent quark model to depict the magnetic moment of hadron, it is natural to extend this model to investigate the magnetic moments of the hadronic molecules. In Refs. \cite{Liu:2003ab,Wang:2016dzu,Gao:2021hmv,Li:2021ryu}, the magnetic moments of  the hadronic molecular states have been studied in the framework of the constituent quark model based on different combinations of the flavour-spin wave functions of the component hadrons. In particular, the authors of Ref. \cite{Li:2021ryu} discussed the magnetic moments and the transition magnetic moments of the observed $P_c/P_{cs}$ states under the hadronic molecular picture through the constituent quark model.

In this work, we mainly focus on the magnetic moments of these $S$-wave $\Sigma_c^{(*)}D^{(*)}$-type doubly charmed molecular pentaquark candidates, which are physics quantities to reflect their inner structures. Nevertheless, these states may have very short lifetime. Thus, it is not an easy task to measure their magnetic moments experimentally. In order to overcome the above difficulty, we further study the transition magnetic moments and the radiative decay behaviors between these possible $S$-wave $\Sigma_c^{(*)}D^{(*)}$-type doubly charmed molecular pentaquarks, which can be regarded as alternative physical observable quantities to test our theoretical prediction. In the realistic calculation, both the $S$-$D$ wave mixing effect and the coupled channel effect are taken into account \cite{Chen:2021kad}. Here, we should mention that the radiative decay processes can be viewed as the ideal platforms to study the electromagnetic properties of these discussed $S$-wave $\Sigma_c^{(*)}D^{(*)}$-type doubly charmed molecular pentaquark candidates (see Fig. \ref{candidates}). Thus, it is highly probable that their electromagnetic properties can be detected in the future experiments, which can also be obtained from the lattice QCD simulations. Hopefully, our study will also inspire experimental colleagues to focus on the electromagnetic properties of the hadronic molecular states in future.

The remainder of this paper is organized as follows. In Sec. \ref{sec2}, we illustrate how to deduce the magnetic moments, the transition magnetic moments, and the radiative decay widths for the hadronic states in the framework of the constituent quark model. With this preparation, we present the numerical results and discussions of the magnetic moments, the transition magnetic moments, and the radiative decay widths for these discussed $S$-wave $\Sigma_c^{(*)}D^{(*)}$-type doubly charmed molecular pentaquark candidates by performing the single channel analysis and the coupled channel analysis in Sec. \ref{sec3}. Finally, we will give a summary in Sec. \ref{sec4}.

\section{The method of calculating the magnetic moments and transition magnetic moments}\label{sec2}

As a well known fact, studying the electromagnetic properties especially including the magnetic moments of hadrons is the crucial aspect to reflect their inner structures. Different theoretical models and approaches have been developed to quantitatively calculate these physics observable quantities, which include the extended Nambu-Jona-Lasinio model, the Bag model, the constituent quark model, the QCD sum rules, the lattice QCD simulations, the chiral perturbation theory, and so on \cite{Meng:2022ozq}. In the present work, we mainly focus on the magnetic moments, the transition magnetic moments, and the radiative decay widths of the $S$-wave $\Sigma_c^{(*)}D^{(*)}$-type doubly charmed molecular pentaquark candidates in the framework of the constituent quark model, which was applied to study the magnetic moments of the hadronic molecular states \cite{Liu:2003ab,Wang:2016dzu,Gao:2021hmv,Li:2021ryu}. In this work, the adopted convention of calculating the magnetic moments and the transition magnetic moments of these discussed $S$-wave $\Sigma_c^{(*)}D^{(*)}$-type doubly charmed molecular pentaquark candidates is the same as that in Ref. \cite{Li:2021ryu}.

In this section, we introduce the method of estimating the magnetic moments and the transition magnetic moments of the $S$-wave $\Sigma_c^{(*)}D^{(*)}$-type doubly charmed molecular pentaquark candidates. In the framework of the constituent quark model, the total magnetic moment $\vec{\mu}$ of a compound system contains the spin magnetic moment $\vec{\mu}_{{\rm spin}}$ and the orbital magnetic moment $\vec{\mu}_{{\rm orbital}}$ from all of its constituents \cite{Huang:2004tn,Wang:2016dzu,Liu:2003ab}, which can be written in a general form, i.e.,
\begin{equation}
 \vec{\mu}=\vec{\mu}_{{\rm spin}}+\vec{\mu}_{{\rm orbital}}. 
\end{equation}

In the above expression, the spin magnetic moment $\vec{\mu}_{{\rm spin}}$ can be explicitly expressed as
\begin{equation}
\vec{\mu}_{{\rm spin}}=\sum_{i}\mu_{i} \overrightarrow{\sigma_{i}},
\end{equation}
where $\mu_i=Q_i/2M_i$ is the magnetic moment of the $i$-th constituent with $Q_i$ and $M_i$ as the charge and mass of the $i$-th constituent, respectively, and $\overrightarrow{\sigma_{i}}$ denotes the Pauli's spin matrix of the $i$-th constituent.

In addition, the explicit expression of the orbital magnetic moment $\vec{\mu}_{{\rm orbital}}$ of the baryon-meson system is given by \cite{Liu:2003ab}
\begin{equation}
  \vec{\mu}_{{\rm orbital}}=\frac{m_{m} \mu_{b}}{m_{b}+m_{m}} \vec{l}+\frac{m_{b} \mu_{m}}{m_{b}+m_{m}} \vec{l}, \label{orbitalmagneticmoment}
\end{equation}
where the subscripts $b$ and $m$ are applied to distinguish these values corresponding to the baryon and meson, respectively, and $\vec{l}$ denotes the orbital angular momenta between the baryon and meson. As shown in Eq. (\ref{orbitalmagneticmoment}), obviously, the contribution of the orbital magnetic moment $\vec{\mu}_{{\rm orbital}}$ of the $S$-wave hadronic state is zero, while there only exists the spin magnetic moment $\vec{\mu}_{{\rm spin}}$. Based on the analysis mentioned above, we can also find that the magnetic moment of hadron is relevant to the spin, flavor, charge, mass, and orbital angular momenta of these involved hadronic states.

In general, there are two crucial steps when extracting the magnetic moment of the discussed hadron:
\begin{itemize}
  \item Firstly, we need to construct the flavor-spin wave function $\phi_{H}$ for the investigated hadron.
  \item Secondly, the magnetic moment of the discussed hadron can be calculated by the $z$-component of the magnetic moment operator $\hat{\mu}^z$ sandwiched by the corresponding flavor-spin wave function $\phi_{H}$ of the investigated hadron, i.e.,
\begin{equation}
\mu_{H}=\left\langle\phi_{H} \left|\hat{\mu}^z \right| \phi_{H}\right\rangle, \label{expectationvalue}
\end{equation}
where $\hat{\mu}^z=\hat{\mu}_{{\rm spin}}^z+\hat{\mu}_{{\rm orbital}}^z$ denotes the $z$-component of the total magnetic moment operator and $\mu_{H}$ is the obtained magnetic moment for the discussed hadron.
\end{itemize}
As a result of these manipulations, we obtain the magnetic moment of the discussed hadron quantitatively.

We should emphasize that the expectation values of the $z$-component of the spin magnetic moment operator $\hat{\mu}_{{\rm spin}}^z$ sandwiched by the corresponding flavor-spin wave functions of quark are
\begin{equation}
  \left\langle q\uparrow \left| \hat{\mu}_{{\rm spin}}^z \right| q\uparrow \right\rangle = + \mu_q,~~~\left\langle q\downarrow  \left| \hat{\mu}_{{\rm spin}}^z \right| q\downarrow  \right\rangle = - \mu_q
\end{equation}
with $\mu_q=-\mu_{\overline{q}}={Q_q}/{2M_q}$, where $M_q$ and $Q_q$ are the quark mass and charge, respectively.

\subsection{Magnetic moments and transition magnetic moments of the $S$-wave charmed baryon $\Sigma_c^{(*)}$ and charmed meson $D^{(*)}$}\label{A}

Within the framework of the constituent quark model, the magnetic moments of the $S$-wave $\Sigma_c^{(*)}D^{(*)}$-type doubly charmed molecular pentaquark candidates can be related to the magnetic moments of the linear combination of the $S$-wave charmed baryon $\Sigma_c^{(*)}$ and the $S$-wave charmed meson $D^{(*)}$. Since the experimental values of the magnetic moments of the $S$-wave charmed baryon $\Sigma_c^{(*)}$ and the $S$-wave charmed meson $D^{(*)}$ are still absent, we directly calculate their magnetic moments by the constituent quark model.

In the following, we take the $S$-wave charmed meson $D^{*0}$ as an example to illustrate how to obtain the magnetic moments of  the conventional mesons and baryons. For the $S$-wave charmed meson $D^{*0}$, the flavor-spin wave function can be written as
\begin{equation}
\phi_{D^{*0}}^{S=1;\,S_3=1}=\left|c\overline{u}\right\rangle\left|\uparrow\uparrow\right\rangle. 
\end{equation}
Here, $\uparrow$ means that the third component of the quark spin is ${1}/{2}$. By Eq. (\ref{expectationvalue}), the magnetic moment of  the $S$-wave charmed meson $D^{*0}$ can be obtained by the $z$-component of the magnetic moment operator $\hat{\mu}^z$ sandwiched by the corresponding flavor-spin wave functions of the $S$-wave charmed meson $D^{*0}$, i.e.,
\begin{eqnarray}
\mu_{D^{*0}}&=&\left\langle \phi_{D^{*0}}^{S=1;\,S_3=1} \left|\hat{\mu}^z\right| \phi_{D^{*0}}^{S=1;\,S_3=1} \right\rangle\nonumber\\
&=& \left\langle c\overline{u}\uparrow\uparrow \left|\hat{\mu}^z\right| c\overline{u}\uparrow\uparrow\right\rangle\nonumber\\
&=&\mu_c+\mu_{\overline{u}}. 
\end{eqnarray}
Thus, the magnetic moment of the $S$-wave charmed meson $D^{*0}$ is $\mu_c+\mu_{\overline{u}}$. Since the charmed meson $D^{*0}$ is the $S$-wave hadron, there only exists the spin magnetic moment, and the contribution of the orbital magnetic moment disappears.

Furthermore, the transition magnetic moments between the $S$-wave $\Sigma_c^{(*)}D^{(*)}$-type doubly charmed molecular pentaquark candidates are connected to the linear combination of the $S$-wave charmed baryon $\Sigma_c^{(*)}$ and the $S$-wave charmed meson $D^{(*)}$ transition magnetic moments. In the following, we discuss the transition magnetic moments of  the $S$-wave charmed baryon $\Sigma_c^{(*)}$ and the $S$-wave charmed meson $D^{(*)}$. Here, the method of getting the transition magnetic moments between the hadrons is similar to that of the magnetic moments of the hadrons, except for the different flavor-spin wave functions of the initial and final states, i.e., $\mu_{H^{\prime} \to H}=\left\langle\phi_{H} \left|\hat{\mu}^z \right| \phi_{H^{\prime}}\right\rangle$.

Here, we discuss the transition magnetic moment of  the $D^{*0}\rightarrow D^0\gamma~$ process. For the $S$-wave charmed mesons $D^{*0}$ and $D^{0}$, we  construct their flavor-spin wave functions as
\begin{eqnarray}
\phi_{D^{*0}}^{S=1;\,S_3=0}=\frac{1}{\sqrt{2}}\left|c\overline{u}\right\rangle\left|\uparrow\downarrow+\downarrow\uparrow\right\rangle,\\
\phi_{D^{0}}^{S=0;\,S_3=0}=\frac{1}{\sqrt{2}}\left|c\overline{u}\right\rangle\left|\uparrow\downarrow-\downarrow\uparrow\right\rangle.
\end{eqnarray}
And then, the transition magnetic momentum of the $D^{*0}\rightarrow D^0\gamma~$ process can be given by the $z$-component of the magnetic moment operator $\hat{\mu}^z$ sandwiched by the flavor-spin wave functions of the $S$-wave charmed mesons $D^{*0}$ and $D^{0}$, i.e.,
\begin{eqnarray}
  \mu_{D^{*0}\rightarrow D^0}&=&\left\langle\phi_{D^{0}}^{S=0;\,S_3=0} \left|\hat{\mu}^z\right| \phi_{D^{*0}}^{S=1;\,S_3=0}\right\rangle \notag\\
  &=& \left\langle \frac{c\overline{u}\uparrow \downarrow-c\overline{u}\downarrow\uparrow}{\sqrt{2}}\right|\hat{\mu}^z\left|\frac{c\overline{u}\uparrow\downarrow+c\overline{u}\downarrow\uparrow}{\sqrt{2}} \right\rangle \notag\\
  &=&\mu_c-\mu_{\overline{u}}. 
\end{eqnarray}
As a result, the transition magnetic moment of the $D^{*0}\to D^0\gamma~$ process denotes $\mu_c-\mu_{\overline{u}}$.

Similarly to the deduced procedure of the magnetic moment of the $S$-wave charmed meson $D^{*0}$ and the transition magnetic moment of the $D^{*0}\to D^0\gamma~$ process, we get the expressions of the magnetic moments and the transition magnetic moments of the $S$-wave charmed baryon $\Sigma_c^{(*)}$ and the $S$-wave charmed meson $D^{(*)}$, which are collected in Tables \ref{Table1} and \ref{Table2}.
\renewcommand\tabcolsep{0.00cm}
\renewcommand{\arraystretch}{1.50}
\begin{table}[!htbp]
  \caption{Magnetic moments of the $S$-wave charmed baryon $\Sigma_c^{(*)}$ and the $S$-wave charmed meson $D^{*}$. Here, the magnetic moment is in unit of the nuclear magnetic moment $\mu_N$, while the magnetic moment of the $S$-wave charmed meson $D$ is zero.}
  \label{Table1}
\begin{tabular}{l|c|ccccc}
\toprule[1.0pt]
\toprule[1.0pt]
\multirow{2}{*}{Hadrons} & \multirow{2}{*}{\textbf{$I(J^P)$}} & \multicolumn{5}{c}{Magnetic moments} \\
   &   & Expressions & Results & \cite{Simonis:2016pnh} & \cite{Ghalenovi:2021jex} & \cite{Zhang:2021mam} \\ \cline{1-7}
$\Sigma^{++}_c$ & $1(\frac{1}{2}^+)$ & $\frac{4}{3}\mu_u-\frac{1}{3}\mu_c$ & $2.357$ &  & $2.450$ & $2.130$ \\
$\Sigma^+_c$ & $ $ & $\frac{2}{3}\mu_u+\frac{2}{3}\mu_d-\frac{1}{3}\mu_c$ & $0.496$ &  & $0.520$ & $0.410$ \\
$\Sigma^0_c$ & $ $ & $\frac{4}{3}\mu_d-\frac{1}{3}\mu_c$ & $-1.365$ &  & $-1.390$ & $-1.310$ \\
$\Sigma^{*++}_c$ & $1(\frac{3}{2}^+)$ & $2\mu_u+\mu_c$ & $4.094$ &  & $4.100$ & $4.070$ \\
$\Sigma^{*+}_c$ & $ $ & $\mu_u+\mu_d+\mu_c$ & $1.302$ &  & $1.270$ & $1.390$ \\
$\Sigma^{*0}_c$ & $ $ & $2\mu_d+\mu_c$ & $-1.490$ &  & $-1.540$ & $-1.290$ \\
$D^{*0}$ & $\frac{1}{2} (1^-)$ & $\mu_c + \mu_ {\overline{u}}$ & $-1.489$ & $-1.470$ &  &  \\
$D^{*+}$ & $ $ & $\mu_c+\mu_{\overline{d}}$ & $1.303$ & $1.320$ &  & $1.210$ \\
\bottomrule[1.0pt]
\bottomrule[1.0pt]
\end{tabular}
\end{table}

\renewcommand\tabcolsep{0.01cm}
\renewcommand{\arraystretch}{1.50}
\begin{table}[!htbp]
  \caption{Transition magnetic moments of the $S$-wave charmed baryon $\Sigma_c^{(*)}$ and the $S$-wave charmed meson $D^{(*)}$. Here, the transition magnetic moment is in unit of the nuclear magnetic moment $\mu_N$.}
  \label{Table2}
\begin{tabular}{l|ccccc}
\toprule[1.0pt]
\toprule[1.0pt]
\multirow{2}{*}{Decay modes} & \multicolumn{5}{c}{Transition magnetic moments} \\
  & Expressions & Results & \cite{Simonis:2016pnh} & \cite{Ghalenovi:2021jex} & \cite{Simonis:2018rld} \\ \cline{1-6}
$\Sigma^{*++}_c \to \Sigma^{++}_c\gamma$ & $\frac{2\sqrt{2}}{3}(\mu_u-\mu_c)$ & $1.404$ &  & $1.470$ & $1.340$\\
$\Sigma^{*+}_c \to \Sigma^+_c\gamma$ & $\frac{\sqrt{2}}{3}(\mu_u+\mu_d-2\mu_c)$ & $0.088$ &  & $0.120$ & $0.102$\\
$\Sigma^{*0}_c \to \Sigma^0_c\gamma$ & $\frac{2\sqrt{2}}{3}(\mu_d-\mu_c)$ & $-1.228$ &  &  & $-1.140$\\
$D^{*0} \to D^0\gamma$ & $\mu_c - \mu_ {\overline{u}}$ & $2.233$ & $2.250$ &  & \\
$D^{*+} \to D^+\gamma$ & $\mu_c-\mu_{\overline{d}}$ & $-0.559$ & $-0.540$ &  & \\
\bottomrule[1.0pt]
\bottomrule[1.0pt]
\end{tabular}
\end{table}

For describing these discussed magnetic moments and transition magnetic moments quantitatively, we take the values of the involved quark masses as input.  In this work,  $m_{u}=0.336~\mathrm{GeV}$, $m_{d}=0.336~\mathrm{GeV}$, and $m_{c}=1.680 ~\mathrm{GeV}$ are adopted \cite{Kumar:2005ei}. In order to check the reliability of the above input parameters, we briefly discuss the magnetic moments of the proton and neutron, since there exists the experimental information of the magnetic moments of the proton and neutron \cite{Workman:2022ynf}. Based on the constituent quark model and the above input parameters, we obtain the magnetic moment of proton $\mu_p=2.793\mu_N$ and the magnetic moment of neutron $\mu_n=-1.862\mu_N$, which are consistent with the experiment values $\mu_p^{\rm{Expt}}=2.793\mu_N$ and $\mu_n^{\rm{Expt}}=-1.913\mu_N$ \cite{Workman:2022ynf}, respectively. Here, $\mu_N={e}/{2M_N}$ denotes the nuclear magnetic moment, while $m_N$ is the nuclear mass with $m_{N}=0.938 ~\mathrm{GeV}$. Thus, we test the reliability of the adopted constituent quark model.

In Tables \ref{Table1} and \ref{Table2}, the calculated magnetic moments and transition magnetic moments of the $S$-wave charmed baryon $\Sigma_c^{(*)}$ and the $S$-wave charmed meson $D^{(*)}$ are collected, which are as input to further calculate the magnetic moments and the transition magnetic moments of  the $S$-wave $\Sigma_c^{(*)}D^{(*)}$-type doubly charmed molecular pentaquark candidates. Here, 
our obtained numerical results are comparable with that from Refs. \cite{Simonis:2016pnh,Ghalenovi:2021jex,Zhang:2021mam,Simonis:2018rld}.

\subsection{Magnetic moments and transition magnetic moments of the $S$-wave $\Sigma_c^{(*)}D^{(*)}$ molecules}\label{B}

In this subsection, we mainly illustrate the method of getting the magnetic moments and the transition magnetic moments of these $S$-wave $\Sigma_c^{(*)}D^{(*)}$-type doubly charmed molecular pentaquark candidates by performing the single channel and coupled channel analysis. In the coupled channel analysis, the relevant channels $|^{2S+1}L_J\rangle$ involved in our calculation are summarized in Table \ref{Table3}.

\renewcommand\tabcolsep{0.32cm}
\renewcommand{\arraystretch}{1.50}
\begin{table}[!htbp]
  \caption{The relevant channels $|^{2S+1}L_J\rangle$ involved in our calculation when considering the coupled channel effect. Here, $S$, $L$, and $J$ denote the spin, relative angular momentum, and total angular momentum of the corresponding channels, respectively.}
  \label{Table3}
\begin{tabular}{l|llll}
\toprule[1.0pt]
\toprule[1.0pt]
$|^{2S+1}L_J\rangle$ & $\Sigma_{c} D$ & $\Sigma_{c}^{*} D$ & $\Sigma_{c} D^{*}$ & $\Sigma_{c}^{*} D^{*}$ \\
\hline
\multirow{3}{*}{$J=1/2$} & $|^{2} S_{1 / 2}\rangle$ & $|^{4} D_{1 / 2}\rangle$ & $|^{2} S_{1 / 2}\rangle$ & $|^{2} S_{1 / 2}\rangle$\\
  &  &  & $|^{4} D_{1 / 2}\rangle$ & $|^{4} D_{1 / 2}\rangle$\\
  & & & & $|^{6} D_{1 / 2}\rangle$ \\
\hline
\multirow{4}{*}{$J=3/2$} &   & $|^{4} S_{3 / 2}\rangle$ & $|^{4} S_{3 / 2}\rangle$ & $|^{4} S_{3 / 2}\rangle$\\
  &  & $|^{4} D_{3 / 2}\rangle$ & $|^{2} D_{3 / 2}\rangle$ & $|^{2} D_{3 / 2}\rangle$\\
  & & & $|^{4} D_{3 / 2}\rangle$ & $|^{4} D_{3 / 2}\rangle$ \\
  & & & & $|^{6} D_{3 / 2}\rangle$ \\
\bottomrule[1.0pt]
\bottomrule[1.0pt]
\end{tabular}
\end{table}

For the $S$-wave $\Sigma_c^{(*)}D^{(*)}$-type doubly charmed molecular pentaquark candidates, their total wave functions $|\psi\rangle$ are compose of the color, spin, flavor, and radial wave functions, which can be expressed as
\begin{equation}
|\psi\rangle=\phi^{\rm{color}}\otimes\phi^{\rm{spin}}\otimes\phi^{\rm{flavor}}\otimes R^{\rm{radial}},
\end{equation}
where $\phi^{\rm{color}}$, $\phi^{\rm{spin}}$, $\phi^{\rm{flavor}}$, and $R^{\rm{radial}}$ denote the color, spin, flavor, and radial wave functions, respectively. The color wave function $\phi^{\rm{color}}$ is simply taken as $1$ for the hadronic molecular state. When discussing the magnetic moment of the investigated hadronic state in the single channel analysis, the overlap of the radial wave function $R^{\rm{radial}}$ is always omitted as customary, since its radial wave function $R^{\rm{radial}}$ satisfies the relation $\langle R^{\rm{radial}} |R^{\rm{radial}}\rangle=1$. However, when studying the magnetic moment of the investigated hadronic state in the coupled channel analysis, we need the radial wave functions $R^{\rm{radial}}$ of these involved channels as the input, and the radial wave functions of the $S$-wave $\Sigma_c^{(*)}D^{(*)}$-type doubly charmed molecular pentaquark candidates can be obtained by the study of their mass spectrum in Ref. \cite{Chen:2021kad}.

As discussed above, it is necessary to construct the flavor and spin wave functions of these discussed $S$-wave $\Sigma_c^{(*)}D^{(*)}$-type doubly charmed molecular pentaquark candidates. In Table \ref{Table34}, the flavor wave functions $\left|I,I_3\right\rangle$ and the spin wave functions $\left|S,S_3\right\rangle$ of the $S$-wave $\Sigma_c^{(*)}D^{(*)}$ systems \cite{Chen:2021kad} are listed.

\renewcommand\tabcolsep{0.11cm}
\renewcommand{\arraystretch}{1.50}
\begin{table}[!htbp]
\caption{The flavor wave functions $\left|I,I_3\right\rangle$ and the spin wave functions $\left|S,S_3\right\rangle$ of the $S$-wave $\Sigma_c^{(*)}D^{(*)}$ systems. Here, $I$ and $I_3$ are the isospin and its third component of the investigated systems, respectively, while $S$ and $S_3$ are the spin and its third component of the investigated systems, respectively.}\label{Table34}
\begin{tabular}{l|l|l}
\toprule[1.0pt]
\toprule[1.0pt]
Systems& $\left|I,I_3\right\rangle$ & Flavor wave functions \\
\hline
\multirow{6}{*}{$\Sigma_c^{(*)}D^{(*)}$}&$\left|\frac{1}{2}, \frac{1}{2}\right\rangle$ & $-\sqrt{\frac{2}{3}}\left|\Sigma_{c}^{(*)++} D^{(*) 0}\right\rangle-\frac{1}{\sqrt{3}}\left|\Sigma_{c}^{(*)+} D^{(*)+}\right\rangle$ \\
&$\left|\frac{1}{2},-\frac{1}{2}\right\rangle$ & $-\frac{1}{\sqrt{3}}\left|\Sigma_{c}^{(*)+} D^{(*) 0}\right\rangle-\sqrt{\frac{2}{3}}\left|\Sigma_{c}^{(*) 0} D^{(*)+}\right\rangle$ \\
&$\left|\frac{3}{2}, \frac{3}{2}\right\rangle$ & $\left|\Sigma_{c}^{(*)++} D^{(*)+}\right\rangle$ \\
&$\left|\frac{3}{2}, \frac{1}{2}\right\rangle$ & $-\frac{1}{\sqrt{3}}\left|\Sigma_{c}^{(*)++} D^{(*) 0}\right\rangle+\sqrt{\frac{2}{3}}\left|\Sigma_{c}^{(*)+} D^{(*)+}\right\rangle$ \\
&$\left|\frac{3}{2},-\frac{1}{2}\right\rangle$ & $-\sqrt{\frac{2}{3}}\left|\Sigma_{c}^{(*)+} D^{(*) 0}\right\rangle+\frac{1}{\sqrt{3}}\left|\Sigma_{c}^{(*) 0} D^{(*)+}\right\rangle$ \\
&$\left|\frac{3}{2},-\frac{3}{2}\right\rangle$ & $-\left|\Sigma_{c}^{(*) 0} D^{(*) 0}\right\rangle$ \\
\hline
Systems&$\left|S,S_3\right\rangle$ & Spin wave functions \\
\hline
\multirow{2}{*}{$\Sigma_cD$}&$\left|\frac{1}{2}, \frac{1}{2}\right\rangle$ & $\left|\frac{1}{2},\frac{1}{2}\right\rangle\left|0,0\right\rangle$ \\
&$\left|\frac{1}{2},-\frac{1}{2}\right\rangle$ & $\left|\frac{1}{2},-\frac{1}{2}\right\rangle\left|0,0\right\rangle$ \\
\multirow{4}{*}{$\Sigma_c^{*}D$}&$\left|\frac{3}{2}, \frac{3}{2}\right\rangle$ & $\left|\frac{3}{2},\frac{3}{2}\right\rangle\left|0,0\right\rangle$ \\
&$\left|\frac{3}{2}, \frac{1}{2}\right\rangle$ & $\left|\frac{3}{2},\frac{1}{2}\right\rangle\left|0,0\right\rangle$ \\
&$\left|\frac{3}{2},-\frac{1}{2}\right\rangle$ & $\left|\frac{3}{2},-\frac{1}{2}\right\rangle\left|0,0\right\rangle$ \\
&$\left|\frac{3}{2},-\frac{3}{2}\right\rangle$ & $\left|\frac{3}{2},-\frac{3}{2}\right\rangle\left|0,0\right\rangle$ \\
\multirow{2}{*}{$\Sigma_cD^{*}$}&$\left|\frac{1}{2}, \frac{1}{2}\right\rangle$ &$\frac{1}{\sqrt{3}}\left|\frac{1}{2},\frac{1}{2}\right\rangle\left|1,0\right\rangle-\sqrt{\frac{2}{3}}\left|\frac{1}{2},-\frac{1}{2}\right\rangle\left|1,1\right\rangle$ \\
&$\left|\frac{1}{2},-\frac{1}{2}\right\rangle$ & $\sqrt{\frac{2}{3}}\left|\frac{1}{2},\frac{1}{2}\right\rangle\left|1,-1\right\rangle-\frac{1}{\sqrt{3}}\left|\frac{1}{2},-\frac{1}{2}\right\rangle\left|1,0\right\rangle$ \\
\multirow{4}{*}{$\Sigma_c^{*}D^{*}$}&$\left|\frac{3}{2}, \frac{3}{2}\right\rangle$ & $\left|\frac{1}{2},\frac{1}{2}\right\rangle\left|1,1\right\rangle$ \\
&$\left|\frac{3}{2}, \frac{1}{2}\right\rangle$ & $\sqrt{\frac{2}{3}}\left|\frac{1}{2},\frac{1}{2}\right\rangle\left|1,0\right\rangle+\frac{1}{\sqrt{3}}\left|\frac{1}{2},-\frac{1}{2}\right\rangle\left|1,1\right\rangle$ \\
&$\left|\frac{3}{2},-\frac{1}{2}\right\rangle$ & $\frac{1}{\sqrt{3}}\left|\frac{1}{2},\frac{1}{2}\right\rangle\left|1,-1\right\rangle+\sqrt{\frac{2}{3}}\left|\frac{1}{2},-\frac{1}{2}\right\rangle\left|1,0\right\rangle$ \\
&$\left|\frac{3}{2},-\frac{3}{2}\right\rangle$ & $\left|\frac{1}{2},-\frac{1}{2}\right\rangle\left|1,-1\right\rangle$ \\
\bottomrule[1.0pt]
\bottomrule[1.0pt]
\end{tabular}
\end{table}

With the above preparation, we further calculate the magnetic moments of these discussed $S$-wave $\Sigma_c^{(*)}D^{(*)}$-type doubly charmed molecular pentaquark candidates by performing the single channel and coupled channel analysis. For the transition magnetic moments between these discussed $S$-wave $\Sigma_c^{(*)}D^{(*)}$-type doubly charmed molecular pentaquark candidates, the calculation method is similar to that of getting the magnetic moments of  these discussed $S$-wave $\Sigma_c^{(*)}D^{(*)}$-type doubly charmed molecular pentaquark candidates, except for the different flavor-spin wave functions of the initial and final states.

In the following, we take the $\Sigma_c D$ state with $I(J^P)={1}/{2}({1}/{2}^-)$ as an example to illustrate how to calculate the magnetic moments of these discussed $S$-wave $\Sigma_c^{(*)}D^{(*)}$-type doubly charmed molecular pentaquark candidates in the single channel and coupled channel analysis. When performing the single channel analysis, the calculation of the magnetic moment of the $\Sigma_c D$ state with $I(J^P)={1}/{2}({1}/{2}^-)$ is quite simple. By Table \ref{Table34}, the flavor-spin wave function of the $\Sigma_c D$ state with $I(J^P)={1}/{2}({1}/{2}^-)$ can be  written as
\begin{eqnarray}
 \phi_{\Sigma_{c} D|^{2} S_{1/2}\rangle} &=&\left[-\sqrt{\frac{2}{3}}\left|\Sigma_{c}^{++} D^{0}\right\rangle-\sqrt{\frac{1}{3}}\left|\Sigma_{c}^{+} D^{+}\right\rangle\right]\nonumber\\&&\otimes \left| \frac{1}{2},\frac{1}{2}\right\rangle \left|0,0\right\rangle,
 \end{eqnarray}
where the third component of the isospin of the $\Sigma_c D$ state with $I(J^P)={1}/{2}({1}/{2}^-)$ is taken as $I_3={1}/{2}$. And then, the magnetic moment of the $\Sigma_c D$ state with $I(J^P)={1}/{2}({1}/{2}^-)$ can be obtained by the following matrix element
\begin{eqnarray}
  \mu_{\Sigma_{c} D|^{2} S_{1/2}\rangle}&=&\left\langle \phi_{\Sigma_{c} D|^{2} S_{1/2}\rangle} \left|\hat{\mu}^z\right| \phi_{\Sigma_{c} D|^{2} S_{1/2}\rangle}\right\rangle\nonumber\\
  &=&\frac{2}{3} \mu_{\Sigma_c^{++}}+\frac{1}{3} \mu_{\Sigma_c^+}.
\end{eqnarray}
Thus, the magnetic moment of the $\Sigma_c D$ state with $I(J^P)={1}/{2}({1}/{2}^-)$ is $\frac{2}{3} \mu_{\Sigma_c^{++}}+\frac{1}{3} \mu_{\Sigma_c^+}$ in the single channel analysis. The orbital magnetic moment is still absent, and there only exists the spin magnetic moment of the pure $S$-wave $\Sigma_c D$ state with $I(J^P)={1}/{2}({1}/{2}^-)$.

Similarly to the study of the mass spectrum of the $S$-wave $\Sigma_c^{(*)}D^{(*)}$-type doubly charmed molecular pentaquark candidates \cite{Chen:2021kad}, we should discuss the roles of the coupled channel effect for the magnetic moments of these discussed $S$-wave $\Sigma_c^{(*)}D^{(*)}$-type doubly charmed molecular pentaquark candidates, which is a lengthy and tedious deduction. When performing the coupled channel analysis for the $\Sigma_c D$ state with $I(J^P)={1}/{2}({1}/{2}^-)$, we  consider the following channels \cite{Chen:2021kad}, i.e.,
\begin{eqnarray*}
&&\Sigma_{c} D|^{2} S_{1/2}\rangle,~\Sigma_{c}^* D|^{4} D_{1/2}\rangle,~\Sigma_{c} D^*|^{2} S_{1/2}\rangle,~\Sigma_{c} D^*|^{4} D_{1/2}\rangle,\nonumber\\
&&\Sigma_{c}^* D^*|^{2} S_{1/2}\rangle,~\Sigma_{c}^* D^*|^{4} D_{1/2}\rangle, ~\Sigma_{c}^* D^*|^{6} D_{1/2}\rangle.
\end{eqnarray*}
Thus, the total magnetic moment of the $\Sigma_c D$ state with $I(J^P)={1}/{2}({1}/{2}^-)$ read as
\begin{eqnarray}
  &&\sum_{i}\mu_{i}\left\langle R_{i}|R_{i}\right\rangle+\sum_{i\neq j} \mu_{i \rightarrow j}\left(\left\langle R_{j}|R_{i}\right\rangle+\left\langle R_{i}|R_{j}\right\rangle\right) \notag\\
  &=&\mu_{S_1}\left\langle R_{S_1}|R_{S_1}\right\rangle+\mu_{S_2}\left\langle R_{S_2}|R_{S_2}\right\rangle+\mu_{S_3} \left\langle R_{S_3}|R_{S_3}\right\rangle \notag\\
  &&+\mu_{D_1}\left\langle R_{D_1}|R_{D_1}\right\rangle+\mu_{D_2}\left\langle R_{D_2}|R_{D_2}\right\rangle+\mu_{D_3}\left\langle R_{D_3}|R_{D_3}\right\rangle \notag\\
  &&+\mu_{D_4}\left\langle R_{D_4}|R_{D_4}\right\rangle+\mu_{S_2 \rightarrow S_1}\left(\left\langle R_{S_1}|R_{S_2}\right\rangle+\left\langle R_{S_2}|R_{S_1}\right\rangle\right)  \notag\\
  &&+\mu_{S_3 \rightarrow S_1}\left(\left\langle R_{S_1}|R_{S_3}\right\rangle+\left\langle R_{S_3}|R_{S_1}\right\rangle\right) \notag\\
  &&+\mu_{S_3 \rightarrow S_2} \left(\left\langle R_{S_2}|R_{S_3}\right\rangle+\left\langle R_{S_3}|R_{S_2}\right\rangle\right) \notag\\
  &&+\mu_{D_2 \rightarrow D_1} \left(\left\langle R_{D_1}|R_{D_2}\right\rangle+\left\langle R_{D_2}|R_{D_1}\right\rangle\right) \notag\\
  &&+\mu_{D_3 \rightarrow D_1} \left(\left\langle R_{D_1}|R_{D_3}\right\rangle+\left\langle R_{D_3}|R_{D_1}\right\rangle\right) \notag\\
  &&+\mu_{D_4 \rightarrow D_1} \left(\left\langle R_{D_1}|R_{D_4}\right\rangle+\left\langle R_{D_4}|R_{D_1}\right\rangle\right) \notag\\
  &&+\mu_{D_3 \rightarrow D_2} \left(\left\langle R_{D_2}|R_{D_3}\right\rangle+\left\langle R_{D_3}|R_{D_2}\right\rangle\right) \notag\\
  &&+\mu_{D_4 \rightarrow D_2} \left(\left\langle R_{D_2}|R_{D_4}\right\rangle+\left\langle R_{D_4}|R_{D_2}\right\rangle\right) \notag\\
  &&+\mu_{D_4 \rightarrow D_3} \left(\left\langle R_{D_3}|R_{D_4}\right\rangle+\left\langle R_{D_4}|R_{D_3}\right\rangle\right). \label{totalmagneticmoment}
\end{eqnarray}
For the convenience of calculation, $\Sigma_{c} D|^{2} S_{1 / 2}\rangle$, $\Sigma_{c} D^*|^{2} S_{1 / 2}\rangle$, $\Sigma_{c}^* D^*|^{2} S_{1 / 2}\rangle$, $\Sigma_{c}^* D|^{4} D_{1 / 2}\rangle$, $\Sigma_{c} D^*|^{4} D_{1 / 2}\rangle$, $\Sigma_{c}^* D^*|^{4} D_{1 / 2}\rangle$, and $\Sigma_{c}^* D^*|^{6} D_{1 / 2}\rangle$ are abbreviated as $S_1$, $S_2$, $S_3$, $D_1$, $D_2$, $D_3$, and $D_4$, respectively. In the above expression, $R_{i}$ denotes the radial wave function of the corresponding $i$-th channel, which can be  obtained directly by the calculation of the mass spectrum of the $S$-wave $\Sigma_c^{(*)}D^{(*)}$ molecules \cite{Chen:2021kad}.

In Eq. (\ref{totalmagneticmoment}), we also need the results of the magnetic moments of the $D$-wave channels and the transition magnetic moments as input, expect for the magnetic moments of  the $S$-wave channels. We introduce how to obtain the magnetic moments of the $D$-wave channels. Since the spin-orbit wave functions of these investigated hadronic states are $\left|{ }^{2 S+1} L_{J}\right\rangle=\sum_{m_{S}, m_{L}} C_{S m_{S}, L m_{L}}^{J, M} \chi_{S, m_{S}}Y_{L, m_{L}}$, the $\left|{ }^{4} D_{1 / 2}\right\rangle$ and $\left|{ }^{6} D_{1 / 2}\right\rangle$ channels can be expanded as
\begin{eqnarray}
\left|{ }^{4} D_{1/2}\right\rangle &=&-\sqrt{\frac{2}{5}}\chi_{{3}/{2},\,-{3}/{2}} Y_{2,\,2}+\sqrt{\frac{3}{10}}\chi_{{3}/{2},\,-{1}/{2}}  Y_{2,\,1}\nonumber\\
  &&-\frac{1}{\sqrt{5}}\chi_{{3}/{2},\,{1}/{2}} Y_{2,\,0}+\frac{1}{\sqrt{10}}\chi_{{3}/{2},\,{3}/{2}} Y_{2,\,-1},\nonumber\\
\left|{ }^{6} D_{1 / 2}\right\rangle&=&\frac{1}{\sqrt{15}}\chi_{{5}/{2},\,-{3}/{2}} Y_{2,\,2}-\sqrt{\frac{2}{15}}\chi_{{5}/{2},\,-{1}/{2}}  Y_{2,\,1}\nonumber\\
  &&+\frac{1}{\sqrt{5}}\chi_{{5}/{2},\,{1}/{2}} Y_{2,\,0}-\frac{2}{\sqrt{15}}\chi_{{5}/{2},\,{3}/{2}} Y_{2,\,-1}\nonumber\\
  &&+\frac{1}{\sqrt{3}}\chi_{{5}/{2},\,{5}/{2}} Y_{2,\,-2}.\label{spinorbitwavefunctionsexpanded}
\end{eqnarray}

Combining Eq. (\ref{spinorbitwavefunctionsexpanded}) with the results of the magnetic moments of the corresponding $S$-wave channels, we further evaluate  the magnetic moments of  the $D$-wave channels. For example, the magnetic moment $\mu_{\Sigma_{c}^* D^*|^{4} D_{1 / 2}\rangle}$ is
\begin{eqnarray}
  \mu_{\Sigma_{c}^* D^*|^{4} D_{1/2}\rangle}&=&\left\langle \phi_{\Sigma_{c}^* D^*(^{4} D_{1/2})} \left|\hat{\mu}^z\right| \phi_{\Sigma_{c}^* D^*(^{4} D_{1/2})}\right\rangle  \nonumber\\
  &=&\frac{2}{5}\left(-\mu_{\Sigma_{c}^* D^*|^{4} S_{3 / 2}\rangle}+2\mu_{\Sigma_{c}^* D^*}^l\right) \nonumber\\
  &&+\frac{3}{10}\left(-\frac{1}{3}\mu_{\Sigma_{c}^* D^*|^{4} S_{3 / 2}\rangle}+\mu_{\Sigma_{c}^* D^*}^l\right)\nonumber\\
  &&+\frac{1}{5}\times \frac{1}{3}\mu_{\Sigma_{c}^* D^*|^{4} S_{3 / 2}\rangle} \nonumber\\
  &&+\frac{1}{10}\left(\mu_{\Sigma_{c}^* D^*|^{4} S_{3 / 2}\rangle}-\mu_{\Sigma_{c}^* D^*}^l\right) \nonumber\\
  &=&-\frac{1}{3}\mu_{\Sigma_{c}^* D^*|^{4} S_{3 / 2}\rangle}+\mu_{\Sigma_{c}^* D^*}^l
\end{eqnarray}
with $\mu_{\Sigma_{c}^* D^*}^l=\frac{m_{D^*} \mu_{\Sigma_{c}^*}+m_{\Sigma_{c}^*} \mu_{D^*}}{m_{\Sigma_{c}^*}+m_{D^*}}$. Thus, we obtain the magnetic moment of the ${\Sigma_{c}^* D^*|^{4} D_{1 / 2}\rangle}$ channel. Here, the total magnetic moment of the $D$-wave hadronic state is obtained by the sum of the spin magnetic moments and the orbital magnetic moments from all of its constituents.

Besides the magnetic moments of the $S$-wave and $D$-wave channels, we also need several transition magnetic moments as input in Eq. (\ref{totalmagneticmoment}). For the sake of simplicity, we take $\mu_{\Sigma_{c} D^*|^{2} S_{1 / 2}\rangle \to \Sigma_{c} D|^{2} S_{1 / 2}\rangle}$ as an example to illustrate the procedure of getting the magnetic moments between the $S$-wave $\Sigma_c^{(*)}D^{(*)}$-type doubly charmed molecular pentaquark candidates. By Table \ref{Table34}, the flavor-spin wave functions of the $\Sigma_{c} D|^{2} S_{1 / 2}\rangle$ and $\Sigma_{c} D^*|^{2} S_{1 / 2}\rangle$ states with $(I,\,I_3)=({1}/{2},\,{1}/{2})$ can be constructed as
\begin{eqnarray}
\phi_{\Sigma_{c} D|^{2} S_{1/2}\rangle}&=&\left[-\sqrt{\frac{2}{3}}\left|\Sigma_{c}^{++} D^{0}\right\rangle-\sqrt{\frac{1}{3}}\left|\Sigma_{c}^{+} D^{+}\right\rangle\right]\otimes
\nonumber\\ 
&&\left| \frac{1}{2},\frac{1}{2}\right\rangle \left|0,0\right\rangle,\nonumber\\
\phi_{\Sigma_{c} D^*|^{2} S_{1 / 2}\rangle}&=&\left[-\sqrt{\frac{2}{3}}\left|\Sigma_{c}^{++} D^{*0}\right\rangle-\sqrt{\frac{1}{3}}\left|\Sigma_{c}^{+} D^{*+}\right\rangle\right]\otimes \nonumber\\ &&\left[\frac{1}{\sqrt{3}}\left|\frac{1}{2},\frac{1}{2}\right\rangle\left|1,0\right\rangle-\sqrt{\frac{2}{3}}\left|\frac{1}{2},-\frac{1}{2}\right\rangle\left|1,1\right\rangle\right],\nonumber\\
\end{eqnarray}
respectively. And then, the transition magnetic moment of the $\Sigma_{c} D^*|^{2} S_{1 / 2}\rangle \to \Sigma_{c} D|^{2} S_{1 / 2}\rangle \gamma$ process can be calculated by the following matrix element
\begin{eqnarray}
  &&\mu_{\Sigma_{c} D^*|^{2} S_{1 / 2}\rangle \to \Sigma_{c} D|^{2} S_{1 / 2}\rangle}\nonumber\\
  &=&\left\langle \phi_{\Sigma_{c} D|^{2} S_{1 / 2}\rangle} \left|\hat{\mu}^z\right| \phi_{\Sigma_{c} D^*|^{2} S_{1 / 2}\rangle}\right\rangle \nonumber\\
  &=&\frac{2}{3\sqrt{3}} \mu_{D^{*0} \to D^0}+\frac{1}{3\sqrt{3}} \mu_{D^{*+} \to D^+}.
\end{eqnarray}
Finally, one obtains the magnetic moment for the $\Sigma_c D$ state with $I(J^P)={1}/{2}({1}/{2}^-)$ by calculating and summing each terms in Eq. (\ref{totalmagneticmoment}).

At the end of this section, we introduce the relation between the radiative decay width and the transition magnetic moment, where the radiative decay width is the observable quantity of the radiative decay processes experimentally. Based on the obtained transition magnetic moment, we can further discuss the radiative decay width, where the radiative decay width $\Gamma\left(H_{1} \to H_{2}\gamma\right)$ can be expressed in terms of the transition magnetic moment $\mu_{H_{1} \to H_{2}}$ as \cite{Franklin:1981rc,Simonis:2018rld}
\begin{equation}
  \Gamma\left(H_{1} \to H_{2}\gamma\right)=\alpha_{\rm {EM}}\frac{\omega^{3}}{M_{P}^{2}} \frac{2}{2J+1} \frac{\mu_{H_{1} \to H_{2}}^{2}}{\mu_N^{2}}, \label{width}
\end{equation}
where $H_{1}$ and $H_{2}$ stand for these discussed $S$-wave $\Sigma_c^{(*)}D^{(*)}$-type doubly charmed molecular pentaquark candidates, $\alpha_{\rm {EM}} \approx {1}/{137}$ is the fine structure constant, $M_P$ denotes the proton mass, $J$ is the total angular momentum of the initial $S$-wave $\Sigma_c^{(*)}D^{(*)}$ molecule, and
\begin{equation}
\omega=\frac{M_1^2-M_2^2}{2M_1} \label{omega}
\end{equation}
is the photon momentum with $M_1$ and $M_2$ as the masses of the initial and final $S$-wave $\Sigma_c^{(*)}D^{(*)}$-type doubly charmed molecular pentaquark candidates, respectively.

\section{Numerical results and discussions}\label{sec3}

In this section, we further present the numerical results and discussions of the magnetic moments, the transition magnetic moments, and the radiative decay widths of the $S$-wave $\Sigma_c^{(*)}D^{(*)}$-type doubly charmed molecular pentaquark candidates. Both the $S$-$D$ wave mixing effect and the coupled channel effect are considered in the realistic calculation. For describing these electromagnetic properties quantitatively, the following parameters involved in the hadron masses,
\begin{eqnarray*}
  &m_{D}=1.867 ~\mathrm{GeV}, \quad &m_{D^*}=2.009 ~\mathrm{GeV}, \\
  &m_{\Sigma_c}=2.454 ~\mathrm{GeV}, \quad &m_{\Sigma_c^*}=2.518 ~\mathrm{GeV}, \label{2.6}
\end{eqnarray*}
are introduced, 
which are taken from the Particle Data Group \cite{Workman:2022ynf}.

With the above preparation, we analyze the magnetic moments, the transition magnetic moments, and the radiative decay widths of the $S$-wave $\Sigma_c^{(*)}D^{(*)}$-type doubly charmed molecular pentaquarks by performing the single channel and coupled channel analysis one by one. Furthermore, we briefly discuss the relations between the magnetic moments and the transition magnetic moments of the $S$-wave $\Sigma_c^{(*)}D^{(*)}$-type doubly charmed molecular pentaquark candidates.

\subsection{The magnetic moments of the $S$-wave $\Sigma_c^{(*)}D^{(*)}$ molecules}\label{A}

\subsubsection{Single channel analysis}\label{1}

In our numerical analysis, we first present the results of the magnetic moments for the $S$-wave $\Sigma_c^{(*)}D^{(*)}$-type doubly charmed molecular pentaquarks by performing the single channel analysis, where the relevant expressions and numerical results are collected in Table \ref{Table4}.
\renewcommand\tabcolsep{0.01cm}
\renewcommand{\arraystretch}{1.50}
\begin{table}[!htbp]
  \caption{Magnetic moments of  the $S$-wave $\Sigma_c^{(*)}D^{(*)}$-type doubly charmed molecular pentaquark candidates obtained by performing the single channel analysis. Here, the magnetic moment is in unit of the nuclear magnetic moment $\mu_N$.}
  \label{Table4}
\begin{tabular}{l|cr|cc}
\toprule[1.0pt]
\toprule[1.0pt]
   States  & $I(J^P)$ & $I_3$ & Expressions & Results \\
\hline
    $\Sigma_c D$ & $\frac{1}{2}(\frac{1}{2}^-)$ & $\frac{1}{2}$ & $\frac{2}{3} \mu_{\Sigma_c^{++}}+\frac{1}{3} \mu_{\Sigma_c^+}$ & $1.737$ \\
    $ $ & $ $ & $-\frac{1}{2}$ & $\frac{2}{3} \mu_{\Sigma_c^0}+\frac{1}{3} \mu_{\Sigma_c^+}$ & $-0.744$\\
    $\Sigma_c^* D$ & $\frac{1}{2}(\frac{3}{2}^-)$ & $\frac{1}{2}$ & $\frac{2}{3} \mu_{\Sigma_c^{*++}}+\frac{1}{3} \mu_{\Sigma_c^{*+}}$ & $3.163$\\
    $ $ & $ $ & $-\frac{1}{2}$ & $\frac{2}{3} \mu_{\Sigma_c^{*0}}+\frac{1}{3} \mu_{\Sigma_c^{*+}}$ & $-0.559$\\
    $\Sigma_c D^*$ & $\frac{1}{2}(\frac{3}{2}^-)$ & $\frac{1}{2}$ & $\frac{2}{3} \mu_{\Sigma_c^{++}}+\frac{1}{3} \mu_{\Sigma_c^+}+\frac{2}{3}\mu _{D^{*0}}+\frac{1}{3}\mu _{D^{*+}}$ & $1.178$ \\
    $ $ & $ $ & $-\frac{1}{2}$ & $\frac{2}{3} \mu_{\Sigma_c^0}+\frac{1}{3} \mu_{\Sigma_c^+}+\frac{2}{3}\mu _{D^{*+}}+\frac{1}{3}\mu _{D^{*0}}$ & $-0.372$ \\
    $\Sigma_c D^*$ & $\frac{1}{2}(\frac{1}{2}^-)$ & $\frac{1}{2}$ & $-\frac{2}{9} \mu_{\Sigma_c^{++}}-\frac{1}{9} \mu_{\Sigma_c^+}+\frac{4}{9}\mu _{D^{*0}}+\frac{2}{9}\mu _{D^{*+}}$ & $-0.951$ \\
    $ $ & $ $ & $-\frac{1}{2}$ & $-\frac{2}{9} \mu_{\Sigma_c^0}-\frac{1}{9} \mu_{\Sigma_c^+}+\frac{4}{9}\mu _{D^{*+}}+\frac{2}{9}\mu _{D^{*0}}$ & $0.496$ \\
    $\Sigma_c D$ & $\frac{3}{2}(\frac{1}{2}^-)$ & $\frac{3}{2}$ & $ \mu_{\Sigma_c^{++}} $ & $2.357$ \\
    $ $ & $ $ & $\frac{1}{2}$ & $\frac{1}{3} \mu_{\Sigma_c^{++}}+\frac{2}{3} \mu_{\Sigma_c^+}$ & $1.116$ \\
    $ $ & $ $ & $-\frac{1}{2}$ & $\frac{1}{3} \mu_{\Sigma_c^0}+\frac{2}{3} \mu_{\Sigma_c^+}$ & $-0.124$ \\
    $ $ & $ $ & $-\frac{3}{2}$ & $ \mu_{\Sigma_c^0} $ & $-1.365$ \\
    $\Sigma_c D^*$ & $\frac{3}{2}(\frac{1}{2}^-)$ & $\frac{3}{2}$ & $-\frac{1}{3} \mu_{\Sigma_c^{++}}+\frac{2}{3}\mu _{D^{*+}}$ & $0.083$ \\
    $ $ & $ $ & $\frac{1}{2}$ & $-\frac{1}{9} \mu_{\Sigma_c^{++}}-\frac{2}{9} \mu_{\Sigma_c^+}+\frac{2}{9}\mu _{D^{*0}}+\frac{4}{9}\mu _{D^{*+}}$ & $-0.124$ \\
    $ $ & $ $ & $-\frac{1}{2}$ & $-\frac{1}{9} \mu_{\Sigma_c^0}-\frac{2}{9} \mu_{\Sigma_c^+}+\frac{2}{9}\mu _{D^{*+}}+\frac{4}{9}\mu _{D^{*0}}$ & $-0.331$ \\
    $ $ & $ $ & $-\frac{3}{2}$ & $-\frac{1}{3} \mu_{\Sigma_c^0}+\frac{2}{3}\mu _{D^{*0}}$ & $-0.538$ \\
\bottomrule[1.0pt]
\bottomrule[1.0pt]
\end{tabular}
\end{table}

According to the numerical results in Table \ref{Table4}, we summarize several key points:
\begin{itemize}
  \item The magnetic moments of the $S$-wave $\Sigma_c^{(*)}D^{(*)}$-type doubly charmed molecular pentaquark candidates vary from $-1.365\mu_N$ to $3.163\mu_N$ for the case of the single channel analysis, where the ${\Sigma_cD|^{2} S_{1/2}\rangle}$ state with $(I,\,I_3)=(3/2,\,-3/2)$ and the ${\Sigma_c^{*}D|^{4} S_{3/2}\rangle}$ state with $(I,\,I_3)=(1/2,\,1/2)$ have the smallest magnetic moment and the largest magnetic moment, respectively.
  
    \item Since the magnetic moment of the $S$-wave charmed meson $D$ is zero, the magnetic moments of the $\Sigma_cD$ state with $I(J^P)=1/2({1/2}^-)$ and the $\Sigma_c^{*}D$ state with $I(J^P)=1/2({3/2}^-)$ only contain the contribution of the magnetic moment of the $S$-wave charmed baryon $\Sigma_c^{(*)}$. Based on flavor wave functions listed in  Table \ref{Table34}, we identify that their $S$-wave charmed baryon compositions are $\Sigma_c^{(*)++}$ and $\Sigma_c^{(*)+}$ with $I_3=1/2$, and $\Sigma_c^{(*)0}$ and $\Sigma_c^{(*)+}$ with $I_3=-1/2$. Since $|\mu_{\Sigma_c^{(*)++}}|$ is greater than $|\mu_{\Sigma_c^{(*)0}}|$, the magnetic moments of the $\Sigma_cD$ state with $I(J^P)=1/2({1/2}^-)$ and the $\Sigma_c^{*}D$ state with $I(J^P)=1/2({3/2}^-)$ satisfy the relation $\left| \mu_{I_3=1/2}\right| >\left|\mu_{I_3=-1/2}\right|$. Similarly to the $\Sigma_cD$ state with $I(J^P)=1/2({1/2}^-)$ and the $\Sigma_c^{*}D$ state with $I(J^P)=1/2({3/2}^-)$, the magnetic moment of the $\Sigma_cD$ state with $I(J^P)=3/2({1/2}^-)$ satisfies the relations $\left| \mu_{I_3=1/2}\right| >\left|\mu_{I_3=-1/2}\right|$ and $\left| \mu_{I_3=3/2}\right| >\left|\mu_{I_3=-3/2}\right|$.
  \item For the $\Sigma_cD^*$ states with $I(J^P)=1/2({1/2}^-,\,{3/2}^-)$ and the $\Sigma_cD^*$ state with $I(J^P)=3/2({1/2}^-)$, their magnetic moments are related to the linear combination of the magnetic moments of  the $S$-wave charmed baryon $\Sigma_c$ and the $S$-wave charmed meson $D^*$.
  \item  The $S$-wave $\Sigma_c^{(*)}D^{(*)}$-type doubly charmed molecular pentaquark candidates with same $I(J^P)$ and different $I_3$ quantum numbers have different magnetic moments, since the $S$-wave $\Sigma_c^{(*)}D^{(*)}$-type doubly charmed molecular pentaquarks with same $I(J^P)$ and different $I_3$ quantum numbers have different flavor wave functions $\left|I,I_3\right\rangle$ as listed in Table \ref{Table34}.
  \item The signs of the magnetic moments of a number of the $S$-wave $\Sigma_c^{(*)}D^{(*)}$-type doubly charmed molecular pentaquarks are the same as that of the proton, while the signs of the magnetic moments of several $S$-wave $\Sigma_c^{(*)}D^{(*)}$-type doubly charmed molecular pentaquarks are same as that of the neutron, which depends on their inner structures. In particular, we should point out that the sign of  the magnetic moment of the $\Sigma_cD^{*}$ state with $I(J^P)=1/2(1/2^-)$ is the opposite to that of the $\Sigma_cD^{*}$ state with $I(J^P)=1/2(3/2^-)$, which is similar to the situation of the magnetic moments of the $P_c(4440)$ and $P_c(4457)$ states \cite{Li:2021ryu}. Thus, measuring the magnetic moments can be applied to distinguish the spin-parity quantum numbers of the $\Sigma_cD^{*}$ system.
\end{itemize}

Inspired by the successful interpretation of the observed three $P_c$ states under the meson-baryon molecular picture \cite{Li:2014gra, Wu:2010jy, Karliner:2015ina, Wang:2011rga, Yang:2011wz, Wu:2012md, Chen:2015loa}, the authors carried out a systematical calculation of the magnetic moments for the $\Sigma_c\bar{D}$ state with $I(J^P)=1/2({1/2}^-)$ and the $\Sigma_c\bar{D}^{*}$ states with $I(J^P)=1/2(1/2^-,\,{3/2}^-)$ in Ref. \cite{Li:2021ryu}. When performing the single channel analysis, the obtained magnetic moments for the $P_c(4312)$, $P_c(4440)$, and $P_c(4457)$ states are
\begin{eqnarray*}
  &&\mu_{P_c(4312)^+}=1.737 \mu_N, ~~~~~~~~\mu_{P_c(4312)^0}=-0.744 \mu_N, \\
  &&\mu_{P_c(4440)^+}=-0.827 \mu_N, ~~~~~\mu_{P_c(4440)^0}=0.620\mu_N, \\
  &&\mu_{P_c(4457)^+}=1.365 \mu_N, ~~~~~~~~\mu_{P_c(4457)^0}=-0.186 \mu_N.
\end{eqnarray*}
By comparing the obtained numerical results, we may find the similarities between the magnetic moments of the $S$-wave $\Sigma_cD^{(*)}$-type doubly charmed molecular pentaquarks and that of the $S$-wave $\Sigma_c\bar D^{(*)}$-type hidden-charm molecular pentaquarks with the same quantum numbers. Firstly, the magnetic moment of the $S$-wave $\Sigma_c D$ state is the same as that of the $S$-wave $\Sigma_c\bar D$ state \cite{Li:2021ryu}, since the magnetic moment of the $S$-wave charmed meson $D$ is zero. Secondly, the magnetic moment signs of the $S$-wave $\Sigma_c D^{(*)}$ states are the same as that of the $S$-wave $\Sigma_c\bar D^{(*)}$ states \cite{Li:2021ryu}, and the numerical results are different slightly, which depend on their flavor-spin wave functions and the magnetic moments of their components.

\begin{figure}[!htbp]
  \centering
  \includegraphics[width=0.47\textwidth]{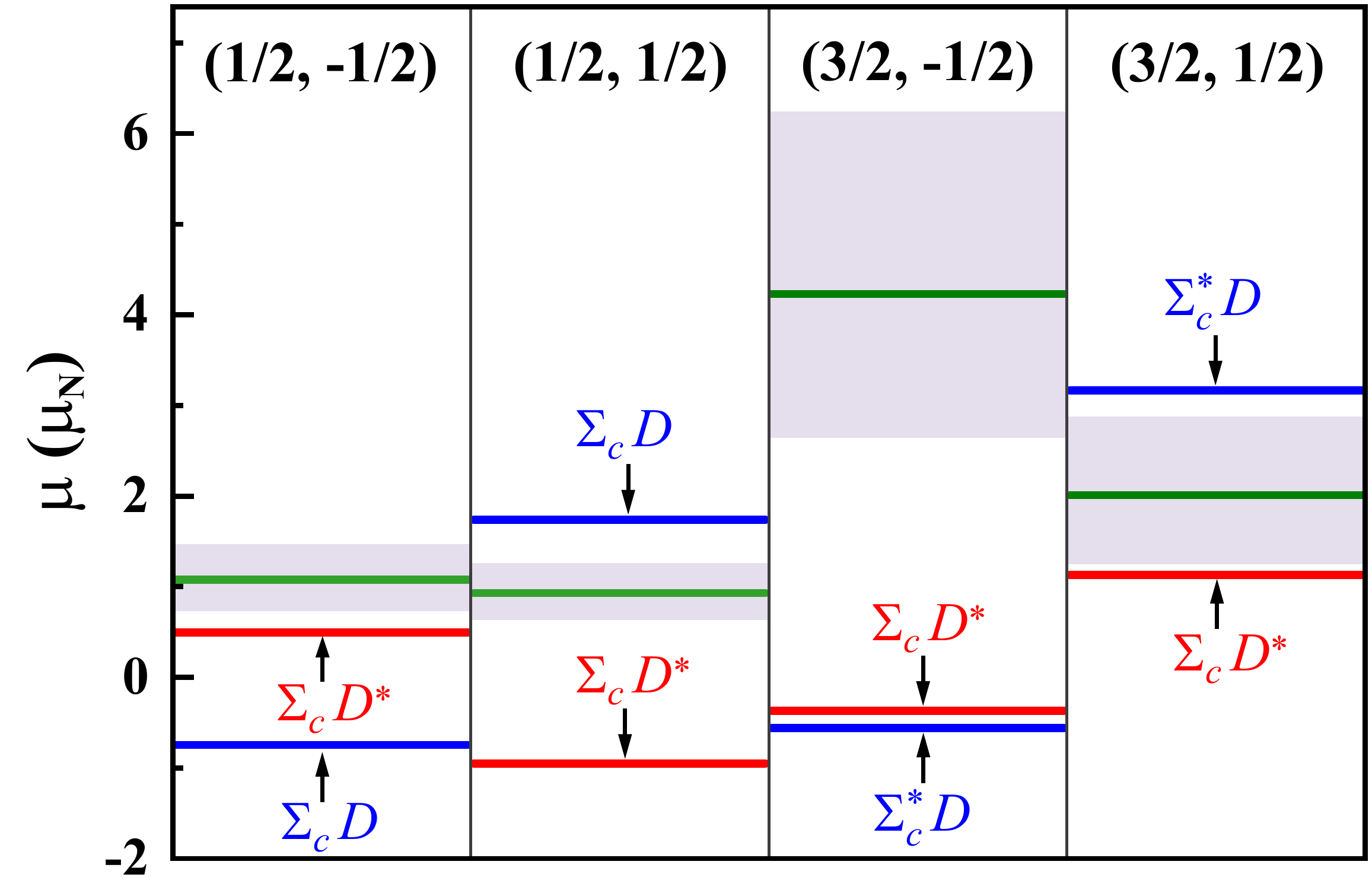}
  \caption{The  comparison of the magnetic moments of the $S$-wave $\Sigma_cD^{(*)}$-type doubly charmed molecular pentaquarks and the compact doubly charmed pentaquarks. The parentheses represent the quantum numbers $(J, I_3)$ for the discussed systems, and their isospin quantum numbers are $1/2$. For the $S$-wave $\Sigma_cD^{(*)}$-type doubly charmed molecular pentaquarks, the red and blue lines stand for the predicted magnetic moments, and we also label corresponding states. For the magnetic moments of  the compact doubly charmed pentaquark states, the green lines denote the central values, while the rectangular areas represent the corresponding errors \cite{Ozdem:2022vip}.}\label{JIzmu}
\end{figure}

As known well, the magnetic moments of the hadronic states usually may reflect their inner structures. In Ref. \cite{Ozdem:2022vip}, the author performed a study of the magnetic moments for the compact doubly charmed pentaquark states with $I(J^P)=1/2(1/2^-,\,{3/2}^-)$ by QCD sum rules. In Fig. \ref{JIzmu}, we compare the magnetic moments of the $S$-wave $\Sigma_cD^{(*)}$-type doubly charmed molecular pentaquarks and the compact doubly charmed pentaquarks with $I(J^P)=1/2(1/2^-,\,{3/2}^-)$. Here, the magnetic moment of the $S$-wave $\Sigma_cD^{(*)}$-type doubly charmed molecular pentaquarks are different from that of the compact doubly charmed pentaquarks with the same quantum number, especially for the doubly charmed pentaquark states with $I_3(J^P)=1/2(1/2^-)$ and $-1/2(3/2^-)$. It shows that the magnetic moment of the hadronic state is one aspect to reflect their inner structures. 

\subsubsection{Coupled channel analysis}\label{2}

After studying the magnetic moments of the $S$-wave $\Sigma_cD^{(*)}$-type doubly charmed molecular pentaquarks by performing the single channel analysis, we further consider the contribution of the coupled channel effect to repeat the numerical analysis of the magnetic moments of the $S$-wave $\Sigma_cD^{(*)}$-type doubly charmed molecular pentaquark candidates.

When calculating the magnetic moments of the $S$-wave $\Sigma_cD^{(*)}$-type doubly charmed molecular pentaquarks by performing the coupled channel analysis, the numerical results of the magnetic moments depend on the relevant channels and the corresponding components, where the components can be obtained from the radial wave functions of the corresponding channels. In our calculations, we take three typical binding energies $-0.5$ MeV, $-6.0$ MeV, and $-13.0$ MeV to show the radial wave functions of the corresponding channels. The magnetic moments of the $S$-wave $\Sigma_c^{(*)}D^{(*)}$-type doubly charmed molecular pentaquark candidates 
are summarized in Table \ref{Table5}.

\renewcommand\tabcolsep{0.15cm}
\renewcommand{\arraystretch}{1.50}
\begin{table}[!htbp]
  \caption{Magnetic moments of  the $S$-wave $\Sigma_c^{(*)}D^{(*)}$-type doubly charmed molecular pentaquark candidates by performing the coupled channel analysis with different binding energies. Here, the magnetic moment is in unit of the nuclear magnetic moment $\mu_N$.}
  \label{Table5}
\begin{tabular}{l|cr|ccc}
\toprule[1.0pt]
\toprule[1.0pt]
  States & $I(J^P)$ & $I_3$ & $-0.5$ MeV & $-6.0$ MeV & $-13.0$ MeV \\
\hline
    $\Sigma_c D$ & $\frac{1}{2}(\frac{1}{2}^-)$ & $\frac{1}{2}$ & $1.731$ & $1.704$ & $1.672$ \\
    $ $ & $ $ & $-\frac{1}{2}$ & $-0.713$ & $-0.633$ & $-0.576$ \\
    $\Sigma_c^* D$ & $\frac{1}{2}(\frac{3}{2}^-)$ & $\frac{1}{2}$ & $3.099$ & $2.854$ & $2.681$ \\
    $ $ & $ $ & $-\frac{1}{2}$ & $-0.549$ & $-0.511$ & $-0.484$ \\
    $\Sigma_c D^*$ & $\frac{1}{2}(\frac{3}{2}^-)$ & $\frac{1}{2}$ & $1.143$ & $1.094$ & $1.072$ \\
    $ $ & $ $ & $-\frac{1}{2}$ & $-0.341$ & $-0.297$ & $-0.278$ \\
    $\Sigma_c D^*$ & $\frac{1}{2}(\frac{1}{2}^-)$ & $\frac{1}{2}$ & $-0.923$ & $-0.875$ & $-0.856$ \\
    $ $ & $ $ & $-\frac{1}{2}$ & $0.496$ & $0.502$ & $0.508$ \\
    $\Sigma_c D$ & $\frac{3}{2}(\frac{1}{2}^-)$ & $\frac{3}{2}$ & $2.307$ & $2.132$ & $1.992$ \\
    $ $ & $ $ & $\frac{1}{2}$ & $1.075$ & $0.960$ & $0.881$ \\
    $ $ & $ $ & $-\frac{1}{2}$ & $-0.163$ & $-0.251$ & $-0.294$ \\
    $ $ & $ $ & $-\frac{3}{2}$ & $-1.403$ & $-1.456$ & $-1.454$ \\
    $\Sigma_c D^*$ & $\frac{3}{2}(\frac{1}{2}^-)$ & $\frac{3}{2}$ & $0.104$ & $0.203$ & $0.339$ \\
    $ $ & $ $ & $\frac{1}{2}$ & $-0.113$ & $-0.058$ & $-0.032$ \\
    $ $ & $ $ & $-\frac{1}{2}$ & $-0.332$ & $-0.337$ & $-0.330$ \\
    $ $ & $ $ & $-\frac{3}{2}$ & $-0.556$ & $-0.621$ & $-0.686$ \\
\bottomrule[1.0pt]
\bottomrule[1.0pt]
\end{tabular}
\end{table}

The results of the magnetic moments of the $S$-wave $\Sigma_cD^{(*)}$-type doubly charmed molecular pentaquarks show:
\begin{enumerate}
  \item We make a comparison of the numerical results from the single channel and coupled channel cases. Obviously, the coupled channel effect can influence the results of the magnetic moments of the $S$-wave $\Sigma_c^{(*)}D^{(*)}$-type doubly charmed molecular pentaquark candidates. In addition, the change of the magnetic moments of the $S$-wave $\Sigma_c^{(*)}D^{(*)}$-type doubly charmed molecular pentaquark candidates is less than $0.48\mu_N$ when adding the contribution of the coupled channel effect. Here, the magnetic moment of the ${\Sigma_c^{*}D|^{4} S_{3/2}\rangle}$ state with $(I,\,I_3)=(1/2,\,1/2)$ is sensitive to the coupled channel effect, which is different from the case for the ${\Sigma_cD^*|^{2} S_{1/2}\rangle}$ state with $(I,\,I_3)=(3/2,\,-1/2)$. These results reflect the dependence of the magnetic moments on the relevant channels and the corresponding components.
  
  \item In the coupled channel analysis, the obtained magnetic moments of the $S$-wave $\Sigma_c^{(*)}D^{(*)}$-type doubly charmed molecular pentaquark candidates are in the range from $-1.456\mu_N$ to $3.099\mu_N$, where the ${\Sigma_cD|^{2} S_{1/2}\rangle}$ state with $(I,\,I_3)=(3/2,\,-3/2)$ and the ${\Sigma_c^{*}D|^{4} S_{3/2}\rangle}$ state with $(I,\,I_3)=(1/2,\,1/2)$ have the smallest magnetic moment and the largest magnetic moment, respectively, which is consistent with the conclusion made by the single channel analysis. Although the magnetic moments depend on the binding energies in the coupled channel analysis, we should point out that the magnetic moments of the $S$-wave $\Sigma_c^{(*)}D^{(*)}$-type doubly charmed molecular pentaquark candidates vary at most $0.42\mu_N$ when we take three typical binding energies $-0.5$ MeV, $-6.0$ MeV, and $-13.0$ MeV.
  
  \item With increasing the binding energies, the magnetic moments of the $S$-wave $\Sigma_c^{(*)}D^{(*)}$-type doubly charmed molecular pentaquark candidates in the coupled channel analysis usually deviate from that in the single channel analysis, since the contributions of the channel with the lowest mass and other channels become unimportant and important, respectively, which can be reflected by the study of the mass spectrum of the $S$-wave $\Sigma_c^{(*)}D^{(*)}$-type doubly charmed molecular pentaquark candidates as given in Ref. \cite{Chen:2021kad}.
\end{enumerate}

\subsection{The transition magnetic moments and radiative decay widths between the $S$-wave $\Sigma_c^{(*)}D^{(*)}$ molecules}\label{B}

In this subsection, we study the transition magnetic moments and the corresponding radiative decay widths between the $S$-wave $\Sigma_c^{(*)}D^{(*)}$-type doubly charmed molecular pentaquark candidates. As shown in Fig. \ref{candidates}, the focused radiative decay processes here include:
(1) $I={1}/{2}$: $\Sigma_c^* D|{3}/{2}^-\rangle \to \Sigma_c D|{1}/{2}^-\rangle \gamma$, $\Sigma_c D^*|{3}/{2}^-\rangle \to \Sigma_c D|{1}/{2}^-\rangle \gamma$, $\Sigma_c D^*|{1}/{2}^-\rangle \to \Sigma_c D|{1}/{2}^-\rangle \gamma$, $\Sigma_c D^*|{3}/{2}^-\rangle \to \Sigma_c^* D|{3}/{2}^-\rangle \gamma $, $\Sigma_c D^*|{1}/{2}^-\rangle \to \Sigma_c^* D|{3}/{2}^-\rangle \gamma$, and $\Sigma_c D^*|{1}/{2}^-\rangle \to \Sigma_c D^*|{3}/{2}^-\rangle \gamma$;
(2) $I={3}/{2}$: $\Sigma_c D^*|{1}/{2}^-\rangle \to \Sigma_c D|{1}/{2}^-\rangle \gamma$.
Since the mass of the $\Sigma_c{D}^{*}$ state with $I(J^P)=1/2({1/2}^-)$ is larger than that of the $\Sigma_c {D}^{*}$ state with $I(J^P)=1/2({3/2}^-)$ when taking the same cutoff parameter \cite{Chen:2021kad}, we only study the $\Sigma_c D^*|{1}/{2}^-\rangle \to \Sigma_c D^*|{3}/{2}^-\rangle \gamma$ process, rather than the $\Sigma_c D^*|{3}/{2}^-\rangle \to \Sigma_c D^*|{1}/{2}^-\rangle \gamma$ process.

\subsubsection{The transition magnetic moments between the $S$-wave $\Sigma_c^{(*)}D^{(*)}$ molecules}\label{1}

We first study the transition magnetic moments between the $S$-wave $\Sigma_c^{(*)}D^{(*)}$-type doubly charmed molecular pentaquark candidates by performing the single channel analysis. In Table \ref{Table6}, the expressions and numerical results of the transition magnetic moments between the $S$-wave $\Sigma_c^{(*)}D^{(*)}$ molecules are listed, where the largest transition magnetic moment corresponds to the $\Sigma_c D^*|{1}/{2}^-\rangle \to \Sigma_c D^*|{3}/{2}^-\rangle\gamma$ process with $(I,\,I_3)=(1/2,\,1/2)$, while the smallest transition magnetic moment is from the $\Sigma_c D^*|{1}/{2}^-\rangle \to \Sigma_c D^*|{3}/{2}^-\rangle\gamma$ process with $(I,\,I_3)=(1/2,\,-1/2)$. Here, the transition magnetic moments of the $\Sigma_c D^*|{3}/{2}^-\rangle \to \Sigma_c^* D|{3}/{2}^-\rangle \gamma $ and $\Sigma_c D^*|{1}/{2}^-\rangle \to \Sigma_c^* D|{3}/{2}^-\rangle \gamma $ processes are zero.

\renewcommand\tabcolsep{0.84cm}
\renewcommand{\arraystretch}{1.50}
\begin{table*}[!htbp]
  \caption{Transition magnetic moments between the $S$-wave $\Sigma_c^{(*)}D^{(*)}$-type doubly charmed molecular pentaquark candidates in the single channel analysis. Here, the transition magnetic moment is in unit of the nuclear magnetic moment $\mu_N$.}
  \label{Table6}
\begin{tabular}{c|r|cc}
\toprule[1.0pt]
\toprule[1.0pt]
Decay modes & $I_3$ & Expressions & Results \\
\hline
$\Sigma_c^* D|\frac{3}{2}^-\rangle \to \Sigma_c D|\frac{1}{2}^-\rangle\gamma$ & $\frac{1}{2}$ & $\frac{2}{3} \mu_{\Sigma_c^{*++} \to \Sigma_c^{++}}+\frac{1}{3} \mu_{\Sigma_c^{*+} \to \Sigma_c^+}$ & $0.965$  \\
$ $ & $-\frac{1}{2}$ & $\frac{2}{3} \mu_{\Sigma_c^{*0} \to \Sigma_c^0}+\frac{1}{3} \mu_{\Sigma_c^{*+} \to \Sigma_c^+}$  & $-0.789$ \\
$\Sigma_c D^*|\frac{3}{2}^-\rangle \to \Sigma_c D|\frac{1}{2}^-\rangle\gamma $ & $\frac{1}{2}$ & $\frac{2}{3}\sqrt{\frac{2}{3}} \mu_{D^{*0} \to D^0}+\frac{1}{3}\sqrt{\frac{2}{3}} \mu_{D^{*+} \to D^+}$ & $1.063$ \\
$ $ & $-\frac{1}{2}$ & $\frac{1}{3}\sqrt{\frac{2}{3}} \mu_{D^{*0} \to D^0}+\frac{2}{3}\sqrt{\frac{2}{3}} \mu_{D^{*+} \to D^+}$  & $0.303$ \\
$\Sigma_c D^*|\frac{1}{2}^-\rangle \to \Sigma_c D|\frac{1}{2}^-\rangle\gamma $ & $\frac{1}{2}$ & $\frac{2}{3\sqrt{3}} \mu_{D^{*0} \to D^0}+\frac{1}{3\sqrt{3}} \mu_{D^{*+} \to D^+}$ & $0.752$ \\
$ $ & $-\frac{1}{2}$ & $\frac{1}{3\sqrt{3}} \mu_{D^{*0} \to D^0}+\frac{2}{3\sqrt{3}} \mu_{D^{*+} \to D^+}$  & $0.215$ \\
$\Sigma_c D^*|\frac{1}{2}^-\rangle \to \Sigma_c D^*|\frac{3}{2}^-\rangle\gamma$ &$\frac{1}{2}$& $\frac{4\sqrt{2}}{9} \mu_{\Sigma_c^{++}}+\frac{2\sqrt{2}}{9} \mu_{\Sigma_c^+}-\frac{2\sqrt{2}}{9}\mu _{D^{*0}}-\frac{\sqrt{2}}{9}\mu _{D^{*+}}$ & $1.901$ \\
 & $-\frac{1}{2}$ & $\frac{4\sqrt{2}}{9} \mu_{\Sigma_c^0}+\frac{2\sqrt{2}}{9} \mu_{\Sigma_c^+}-\frac{2\sqrt{2}}{9}\mu _{D^{*+}}-\frac{\sqrt{2}}{9}\mu _{D^{*0}}$ & $-0.878$ \\
$\Sigma_c D^*|\frac{1}{2}^-\rangle \to \Sigma_c D|\frac{1}{2}^-\rangle\gamma $ & $\frac{3}{2}$ & $\frac{1}{\sqrt{3}} \mu_{D^{*+} \to D^+}$ & $-0.323$ \\
$ $ & $\frac{1}{2}$ & $\frac{1}{3\sqrt{3}} \mu_{D^{*0} \to D^0}+\frac{2}{3\sqrt{3}} \mu_{D^{*+} \to D^+}$ & $0.215$ \\
$ $ & $-\frac{1}{2}$ & $\frac{2}{3\sqrt{3}} \mu_{D^{*0} \to D^0}+\frac{1}{3\sqrt{3}} \mu_{D^{*+} \to D^+}$ & $0.752$ \\
$ $ & $-\frac{3}{2}$ & $\frac{1}{\sqrt{3}} \mu_{D^{*0} \to D^0}$ & $1.289$ \\
\bottomrule[1.0pt]
\bottomrule[1.0pt]
\end{tabular}
\end{table*}

The transition magnetic moments between the hadronic states depend on the flavor-spin wave functions of the initial and final states, and the magnetic moments of their components. As shown in Table \ref{Table6}, the transition magnetic moment of the $\Sigma_c^* D|{3}/{2}^-\rangle \to \Sigma_c D|{1}/{2}^-\rangle\gamma$ process is related to the transition magnetic moment of the $\Sigma_c^* \to \Sigma_c\gamma$ process. The transition magnetic moments of the $\Sigma_c D^*|{3}/{2}^-\rangle \to \Sigma_c D|{1}/{2}^-\rangle\gamma$ and $\Sigma_c D^*|{1}/{2}^-\rangle \to \Sigma_c D|{1}/{2}^-\rangle\gamma$ processes have relations with the transition magnetic moment of the $D^{*} \to D$ process. The transition magnetic moment of the $\Sigma_c D^*|{1}/{2}^-\rangle \to \Sigma_c D^*|{3}/{2}^-\rangle\gamma$ process is connected to the linear combination of the magnetic moments of the $S$-wave charmed baryon $\Sigma_c$ and the $S$-wave charmed meson $D^*$.

Similar to the study of the magnetic moments of the $S$-wave $\Sigma_c^{(*)}D^{(*)}$-type doubly charmed molecular pentaquark candidates, we also further consider the coupled channel effect, and repeat the numerical analysis of the transition magnetic moments between the $S$-wave $\Sigma_c^{(*)}D^{(*)}$-type doubly charmed molecular pentaquark candidates. In Table \ref{Table7}, the transition magnetic moments between the $S$-wave $\Sigma_c^{(*)}D^{(*)}$-type doubly charmed molecular pentaquark candidates are listed. For simplicity, we adopt the same binding energies for the initial and final states and take three typical binding energies $-0.5$ MeV, $-6.0$ MeV, and $-13.0$ MeV to present the transition magnetic moments between the $S$-wave $\Sigma_c^{(*)}D^{(*)}$-type doubly charmed molecular pentaquarks. Our results show  that the largest transition magnetic moment is from the $\Sigma_c D^*|{1}/{2}^-\rangle \to \Sigma_c D^*|{3}/{2}^-\rangle\gamma$ process with $(I,\,I_3)=(1/2,\,1/2)$ and the smallest transition magnetic moment is due to the $\Sigma_c D^*|{1}/{2}^-\rangle \to \Sigma_c D^*|{3}/{2}^-\rangle\gamma$ process with $(I,\,I_3)=(1/2,\,-1/2)$, which is same as the conclusion from the single channel analysis.

\renewcommand\tabcolsep{0.00cm}
\renewcommand{\arraystretch}{1.50}
\begin{table}[!htbp]
  \caption{Transition magnetic moments between the $S$-wave $\Sigma_c^{(*)}D^{(*)}$-type doubly charmed molecular pentaquark candidates by performing the coupled channel analysis with different binding energies. Here, the transition magnetic moment is in unit of the nuclear magnetic moment $\mu_N$.}
  \label{Table7}
\begin{tabular}{c|r|ccc}
\toprule[1.0pt]
\toprule[1.0pt]
Decay modes & $I_3$ & $-0.5$ MeV & $-6.0$ MeV & $-13.0$ MeV \\
\hline
$\Sigma_c^* D|\frac{3}{2}^-\rangle \to \Sigma_c D|\frac{1}{2}^-\rangle\gamma $ & $\frac{1}{2}$ & $1.050$ & $1.265$ & $1.383$ \\
$ $ & $-\frac{1}{2}$ & $-0.764$ & $-0.695$ & $-0.658$ \\
$\Sigma_c D^*|\frac{3}{2}^-\rangle \to \Sigma_c D|\frac{1}{2}^-\rangle\gamma $ & $\frac{1}{2}$ & $1.118$ & $1.257$ & $1.340$ \\
$ $ & $-\frac{1}{2}$ & $0.260$ & $0.164$ & $0.105$ \\
$\Sigma_c D^*|\frac{1}{2}^-\rangle \to \Sigma_c D|\frac{1}{2}^-\rangle\gamma $ & $\frac{1}{2}$ & $0.682$ & $0.521$ & $0.427$ \\
$ $ & $-\frac{1}{2}$ & $0.238$ & $0.300$ & $0.340$ \\
$\Sigma_c D^*|\frac{1}{2}^-\rangle \to \Sigma_c D^*|\frac{3}{2}^-\rangle\gamma $ & $\frac{1}{2}$ & $1.853$ & $1.791$ & $1.762$ \\
$ $ & $-\frac{1}{2}$ & $-0.863$ & $-0.847$ & $-0.838$ \\
$\Sigma_c D^*|\frac{1}{2}^-\rangle \to \Sigma_c D|\frac{1}{2}^-\rangle\gamma $ & $\frac{3}{2}$ & $-0.292$ & $-0.200$ & $-0.127$ \\
$ $ & $\frac{1}{2}$ & $0.233$ & $0.291$ & $0.341$ \\
$ $ & $-\frac{1}{2}$ & $0.754$ & $0.755$ & $0.752$ \\
$ $ & $-\frac{3}{2}$ & $1.276$ & $1.228$ & $1.183$ \\
$\Sigma_c D^*|\frac{3}{2}^-\rangle \to \Sigma_c^* D|\frac{3}{2}^-\rangle\gamma$ & $\frac{1}{2}$ & $-0.103$ & $-0.345$ & $-0.463$ \\
$ $ & $-\frac{1}{2}$ & $0.013$ & $0.045$ & $0.061$ \\
$\Sigma_c D^*|\frac{1}{2}^-\rangle \to \Sigma_c^* D|\frac{3}{2}^-\rangle\gamma$ & $\frac{1}{2}$ & $0.143$ & $0.521$ & $0.715$ \\
$ $ & $-\frac{1}{2}$ & $-0.070$ & $-0.261$ & $-0.362$ \\
\bottomrule[1.0pt]
\bottomrule[1.0pt]
\end{tabular}
\end{table}

For the transition magnetic moments between the $S$-wave hadronic states when considering the coupled channel effect, the transition magnetic moment between the $S$-wave lowest mass channels plays the leading role, and the transition magnetic moments between other channels are subleading. Such contributions from the transition magnetic moments between other channels are usually sensitive to the binding energies. As presented in Table \ref{Table7}, the transition magnetic moments 
of the $\Sigma_c D^*|{3}/{2}^-\rangle \to \Sigma_c^* D|{3}/{2}^-\rangle\gamma$ and $\Sigma_c D^*|{1}/{2}^-\rangle \to \Sigma_c^* D|{3}/{2}^-\rangle\gamma$ processes are sensitive to the binding energies, which is  different from these transition magnetic moments between the $S$-wave $\Sigma_c^{(*)}D^{(*)}$-type doubly charmed molecular pentaquarks. The main reason is that the transition magnetic moments between the $S$-wave lowest mass channels for both processes are zero, and only the transition magnetic moments between other channels contribute to their transition magnetic moments, which makes the numerical results more sensitive to the binding energies.

\subsubsection{The radiative decay widths between the $S$-wave $\Sigma_c^{(*)}D^{(*)}$ molecules}\label{1}

\begin{figure}[!htbp]
  \centering
  \includegraphics[width=0.48\textwidth]{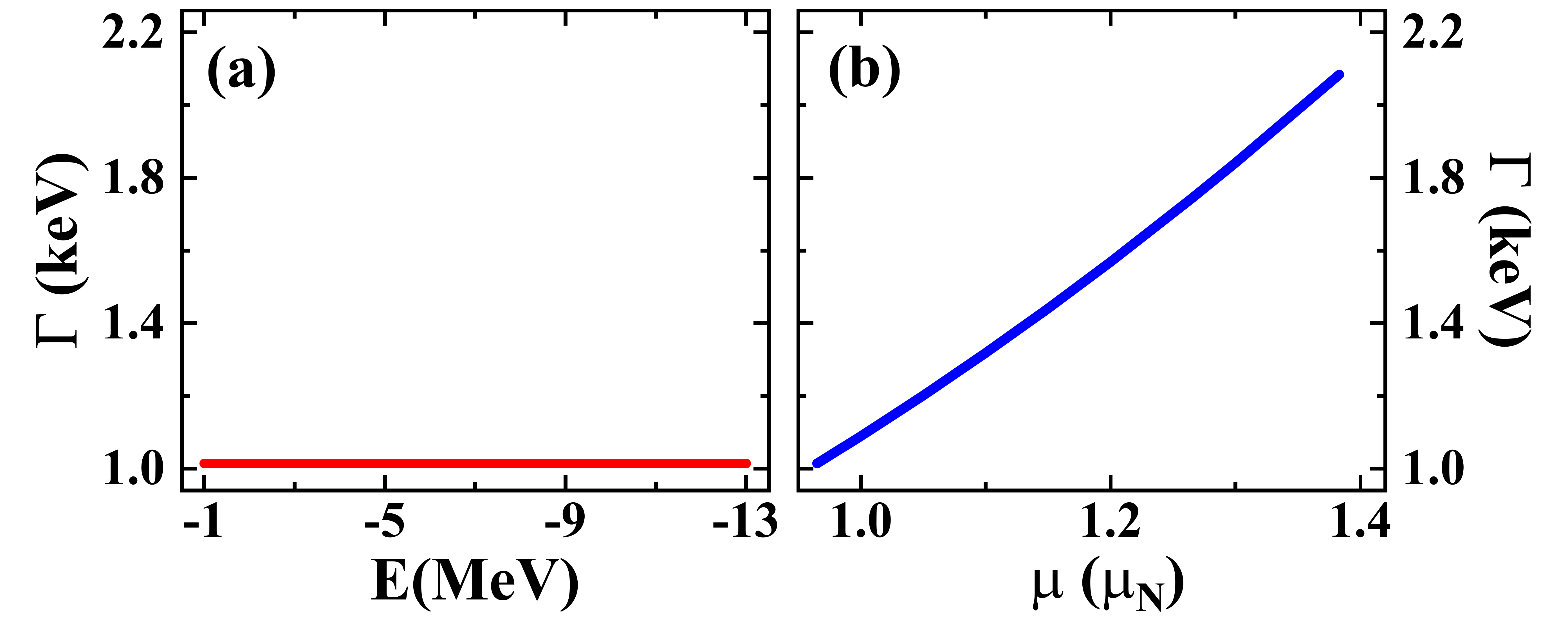}
  \caption{(a) The dependence of the radiative decay width of the $\Sigma_c^* D|{3}/{2}^-\rangle \to \Sigma_c D|{1}/{2}^-\rangle \gamma$ process with $(I,\,I_3)=(1/2,\,1/2)$ on the binding energy when the transition magnetic moment is taken around $0.965\mu_N$. (b) The dependence of the radiative decay width for the $\Sigma_c^* D|{3}/{2}^-\rangle \to \Sigma_c D|{1}/{2}^-\rangle \gamma$ process with $(I,\,I_3)=(1/2,\,1/2)$ on the transition magnetic moment with the binding energy $E=-6.0~{\rm MeV}$. Here, $\Gamma$ is the radiative decay width for the $\Sigma_c^* D|{3}/{2}^-\rangle \to \Sigma_c D|{1}/{2}^-\rangle \gamma$ process with $(I,\,I_3)=(1/2,\,1/2)$, $E$ is the binding energy for the $\Sigma_c^* D$ state with $I(J^P)=1/2({3/2}^-)$ and the $\Sigma_c D$ state with $I(J^P)=1/2({1/2}^-)$, and $\mu$ is the transition magnetic moment of the $\Sigma_c^* D|{3}/{2}^-\rangle \to \Sigma_c D|{1}/{2}^-\rangle \gamma$ process with $(I,\,I_3)=(1/2,\,1/2)$. }\label{EmuGamma}
\end{figure}

Based on the obtained transition magnetic moments between the $S$-wave $\Sigma_c^{(*)}D^{(*)}$ molecules, we can further discuss the corresponding radiative decay widths between the $S$-wave $\Sigma_c^{(*)}D^{(*)}$-type doubly charmed molecular pentaquark candidates. In the following, we take the $\Sigma_c^* D|{3}/{2}^-\rangle \to \Sigma_c D|{1}/{2}^-\rangle \gamma$ process with $(I,\,I_3)=(1/2,\,1/2)$ as an example to illustrate the dependence of the radiative decay width on the binding energy and the transition magnetic moment. In Fig. \ref{EmuGamma}, the dependence of the radiative decay width of the $\Sigma_c^* D|{3}/{2}^-\rangle \to \Sigma_c D|{1}/{2}^-\rangle \gamma$ process with $(I,\,I_3)=(1/2,\,1/2)$ on the binding energy and the transition magnetic moment is given. Here, the binding energies rarely affect the radiative decay width, while the transition magnetic moments obviously change the radiative decay width. Thus, one takes the binding energy as $-6.0$ MeV to present the radiative decay widths between the $S$-wave $\Sigma_c^{(*)}D^{(*)}$-type doubly charmed molecular pentaquark candidates in the single channel analysis.

In Table \ref{Table8}, we collect the radiative decay widths between the $S$-wave $\Sigma_c^{(*)}D^{(*)}$-type doubly charmed molecular pentaquark candidates by performing the single channel and coupled channel analysis. Here, the radiative decay width of the $\Sigma_c D^*|{1}/{2}^-\rangle \to \Sigma_c D^*|{3}/{2}^-\rangle \gamma$ process is zero if  taking the same binding energies for the $\Sigma_c{D}^{*}$ state with $I(J^P)=1/2({1/2}^-)$ and the $\Sigma_c{D}^{*}$ state with $I(J^P)=1/2({3/2}^-)$. In the following analysis, we take different binding energies for the $\Sigma_c{D}^{*}$ state with $I(J^P)=1/2({1/2}^-)$ and the $\Sigma_c{D}^{*}$ state with $I(J^P)=1/2({3/2}^-)$ to discuss the radiative decay width of the $\Sigma_c D^*|{1}/{2}^-\rangle \to \Sigma_c D^*|{3}/{2}^-\rangle \gamma$ process, where the radiative decay width of the $\Sigma_c D^*|{1}/{2}^-\rangle \to \Sigma_c D^*|{3}/{2}^-\rangle \gamma$ process may be strongly suppressed since the mass of the $\Sigma_c{D}^{*}$ state with $I(J^P)=1/2({1/2}^-)$ is extremely close to that of the $\Sigma_c{D}^{*}$ state with $I(J^P)=1/2({3/2}^-)$ in the molecular picture.

\renewcommand\tabcolsep{0.70cm}
\renewcommand{\arraystretch}{1.50}
\begin{table*}[!htbp]
  \caption{The discussion of the radiative decay widths between the $S$-wave $\Sigma_c^{(*)}D^{(*)}$-type doubly charmed molecular pentaquark candidates with two different scenarios: ($\mathcal{A}$) the single channel analysis and ($\mathcal{B}$) the coupled channel analysis. Here, the radiative decay width is in unit of keV.}
  \label{Table8}
\begin{tabular}{c|r|c|ccc}
\toprule[1.0pt]
\toprule[1.0pt]
\multirow{2}{*}{Decay modes} & \multirow{2}{*}{$I_3$} & \multirow{2}{*}{$\mathcal{A}$} &\multicolumn{3}{c}{$\mathcal{B}$} \\
 & & & $-0.5$ MeV & $-6.0$ MeV & $-13.0$ MeV \\
\cline{1-6}
$\Sigma_c^* D|\frac{3}{2}^-\rangle \to \Sigma_c D|\frac{1}{2}^-\rangle\gamma $ & $\frac{1}{2}$ & $1.015$ & $1.201$ & $1.740$ & $ 2.084$ \\
$ $ & $-\frac{1}{2}$ & $0.678$ & $0.636$ & $0.526$ & $0.472$ \\
$\Sigma_c D^*|\frac{3}{2}^-\rangle \to \Sigma_c D|\frac{1}{2}^-\rangle\gamma $ & $\frac{1}{2}$ & $12.602$ & $13.941$ & $17.622$ & $20.024$ \\
$ $ & $-\frac{1}{2}$ & $1.024$ & $0.754$ & $0.300$ & $0.123$ \\
$\Sigma_c D^*|\frac{1}{2}^-\rangle \to \Sigma_c D|\frac{1}{2}^-\rangle\gamma $ & $\frac{1}{2}$ & $12.614$ & $10.375$ & $6.055$ & $4.067$ \\
$ $ & $-\frac{1}{2}$ & $1.031$ & $1.264$ & $2.007$ & $2.578$ \\
$\Sigma_c D^*|\frac{1}{2}^-\rangle \to \Sigma_c D|\frac{1}{2}^-\rangle\gamma $ & $\frac{3}{2}$  & $2.327$ & $1.902$ & $0.892$ & $0.360$ \\
$ $ & $\frac{1}{2}$ & $1.031$ & $1.211$ & $1.889$ & $2.593$ \\
$ $ & $-\frac{1}{2}$ & $12.614$ & $12.682$ & $12.715$ & $12.613$ \\
$ $ & $-\frac{3}{2}$ & $37.061$ & $36.319$ & $33.636$ & $31.214$ \\
$\Sigma_c D^*|\frac{3}{2}^-\rangle \to \Sigma_c^* D|\frac{3}{2}^-\rangle\gamma $ & $\frac{1}{2}$ & $0$ & $0.019$ & $0.218$ & $0.392$ \\
$ $ & $-\frac{1}{2}$ & $0$ & $0$ & $0.004$ & $0.007$ \\
$\Sigma_c D^*|\frac{1}{2}^-\rangle \to \Sigma_c^* D|\frac{3}{2}^-\rangle\gamma $ & $\frac{1}{2}$ & $0$ & $0.075$ & $0.993$ & $1.870$ \\
$ $ & $-\frac{1}{2}$ & $0$ & $0.018$ & $0.249$ & $0.479$ \\
\bottomrule[1.0pt]
\bottomrule[1.0pt]
\end{tabular}
\end{table*}

As shown in Table \ref{Table8}, several radiative decay modes with the decay widths larger than $10.00~{\rm keV}$ are the $\Sigma_c D^*|{3}/{2}^-\rangle \to \Sigma_c D|{1}/{2}^-\rangle\gamma$ process with $(I,\,I_3)=(1/2,\,1/2)$, the $\Sigma_c D^*|{1}/{2}^-\rangle \to \Sigma_c D|{1}/{2}^-\rangle\gamma $ process with $(I,\,I_3)=(1/2,\,1/2)$, the $\Sigma_c D^*|{1}/{2}^-\rangle \to \Sigma_c D|{1}/{2}^-\rangle\gamma $ process with $(I,\,I_3)=(3/2,\,-1/2)$, and the $\Sigma_c D^*|{1}/{2}^-\rangle \to \Sigma_c D|{1}/{2}^-\rangle\gamma $ process with $(I,\,I_3)=(3/2,\,-3/2)$. And, the radiative decay channels with widths smaller than $0.50~{\rm keV}$ include the $\Sigma_c D^*|{3}/{2}^-\rangle \to \Sigma_c^* D|{3}/{2}^-\rangle\gamma$ process with $(I,\,I_3)=(1/2,\,1/2)$, the $\Sigma_c D^*|{3}/{2}^-\rangle \to \Sigma_c^* D|{3}/{2}^-\rangle\gamma$ process with $(I,\,I_3)=(1/2,\,-1/2)$, and the $\Sigma_c D^*|{1}/{2}^-\rangle \to \Sigma_c^* D|{3}/{2}^-\rangle\gamma $ process with $(I,\,I_3)=(1/2,\,-1/2)$. In addition, most of the radiative decay widths between the $S$-wave $\Sigma_c^{(*)}D^{(*)}$-type doubly charmed molecular pentaquark candidates are around $1.00~{\rm keV}$. Furthermore, the coupled channel effect may play an important role to mediate several radiative decay widths between the $S$-wave $\Sigma_c^{(*)}D^{(*)}$-type doubly charmed molecular pentaquark candidates, which is similar to the situation of the radiative decay behaviors between the $S$-wave $\Sigma_c\bar D^{(*)}$-type hidden-charm molecular pentaquarks  \cite{Li:2021ryu}.

Different from other discussed radiative decay processes between the $S$-wave $\Sigma_c^{(*)}D^{(*)}$-type doubly charmed molecular pentaquark candidates, the radiative decay widths of the $\Sigma_c D^*|{3}/{2}^-\rangle \to \Sigma_c^* D|{3}/{2}^-\rangle\gamma$ and $\Sigma_c D^*|{1}/{2}^-\rangle \to \Sigma_c^* D|{3}/{2}^-\rangle\gamma$ processes are sensitive to the binding energies, since the transition magnetic moments of the $\Sigma_c D^*|{3}/{2}^-\rangle \to \Sigma_c^* D|{3}/{2}^-\rangle\gamma$ and $\Sigma_c D^*|{1}/{2}^-\rangle \to \Sigma_c^* D|{3}/{2}^-\rangle\gamma$ processes obviously depend on the binding energies and the radiative decay width is proportional to the transition magnetic moment squared. Thus, the experimental measurement of the binding energies of the relevant doubly charmed molecular pentaquark states will help us to improve our analysis about the radiative decay widths of the $\Sigma_c D^*|{3}/{2}^-\rangle \to \Sigma_c^* D|{3}/{2}^-\rangle\gamma$ and $\Sigma_c D^*|{1}/{2}^-\rangle \to \Sigma_c^* D|{3}/{2}^-\rangle\gamma$ processes \cite{Wang:2019spc,Chen:2017xat,Chen:2021tip}.

As shown in Eq. (\ref{width}), the radiative decay widths between the $S$-wave $\Sigma_c^{(*)}D^{(*)}$-type doubly charmed molecular pentaquark candidates depend on the kinetic phase space and the transition magnetic moment. Here, we take two examples to illustrate this point. Firstly,  the $\Sigma_c^* D|{3}/{2}^-\rangle \to \Sigma_c D|{1}/{2}^-\rangle\gamma $ process with $(I,\,I_3)=(1/2,\,1/2)$ has
very small decay width 
compared with the $\Sigma_c D^*|{3}/{2}^-\rangle \to \Sigma_c D|{1}/{2}^-\rangle\gamma$ process with $(I,\,I_3)=(1/2,\,1/2)$, which is due to the kinetic phase space being suppressed. Secondly, the $\Sigma_c^* D|{3}/{2}^-\rangle \to \Sigma_c D|{1}/{2}^-\rangle\gamma$ process with $(I,\,I_3)=(1/2,\,1/2)$ has decay width larger than that of the $\Sigma_c^* D|{3}/{2}^-\rangle \to \Sigma_c D|{1}/{2}^-\rangle\gamma$ process with $(I,\,I_3)=(1/2,\,-1/2)$, since the transition magnetic moment for the $\Sigma_c^* D|{3}/{2}^-\rangle \to \Sigma_c D|{1}/{2}^-\rangle\gamma$ process with $(I,\,I_3)=(1/2,\,1/2)$ is larger than that of the $\Sigma_c^* D|{3}/{2}^-\rangle \to \Sigma_c D|{1}/{2}^-\rangle\gamma$ process with $(I,\,I_3)=(1/2,\,-1/2)$.

\begin{figure}[!htbp]
  \includegraphics[width=0.45\textwidth]{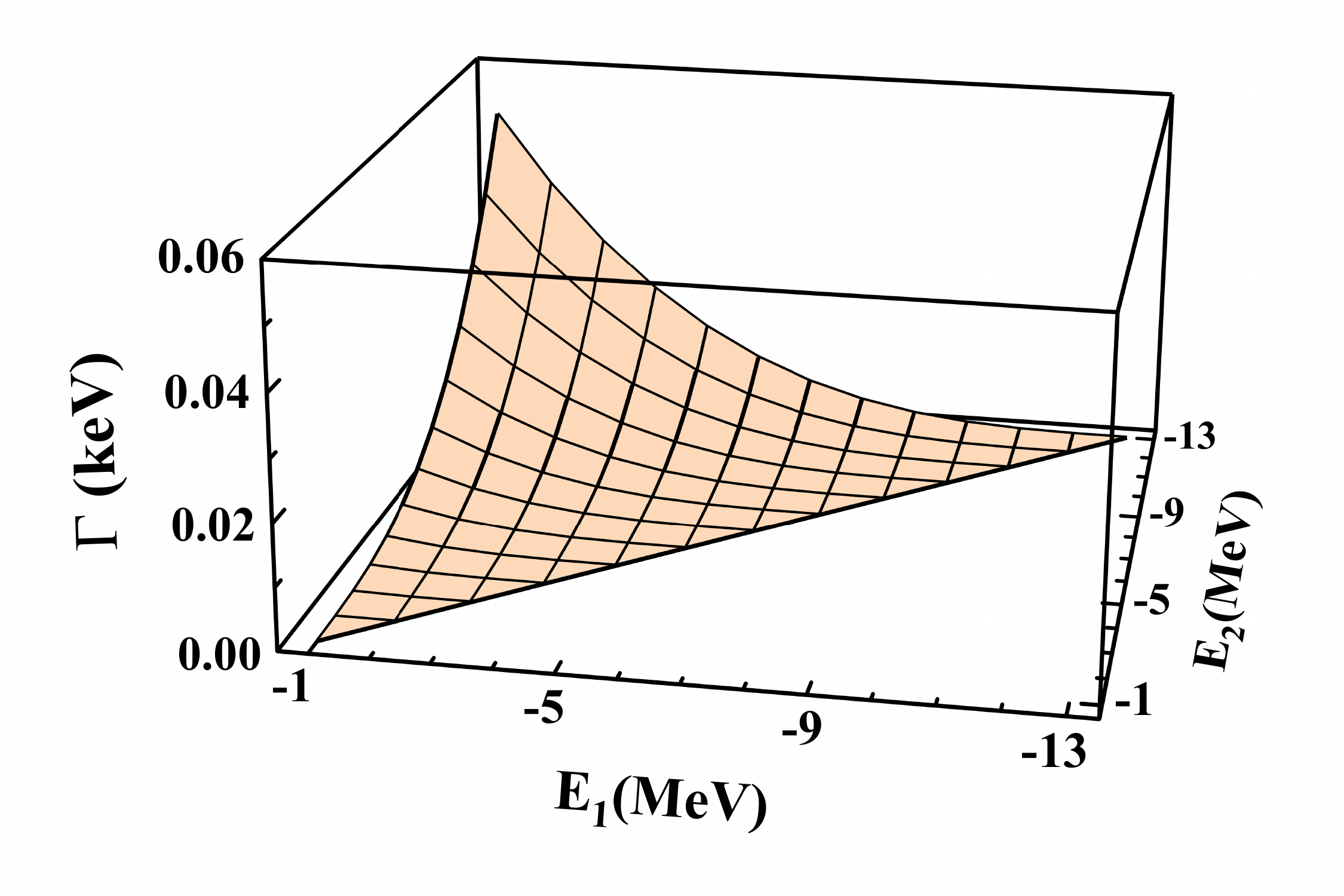}
  \caption{The dependence of the radiative decay width of the $\Sigma_c D^*|{1}/{2}({1}/{2}^-)\rangle\to \Sigma_c D^*|{1}/{2}({3}/{2}^-)\rangle\gamma$ process with $(I,\,I_3)=(1/2,\,1/2)$ on the binding energies of the $\Sigma_c{D}^{*}$ state with $I(J^P)=1/2({1/2}^-)$ and the $\Sigma_c{D}^{*}$ state with $I(J^P)=1/2({3/2}^-)$ when the transition magnetic moment is taken as $1.901\mu_N$. Here, $E_1$ is the binding energy of the $\Sigma_c{D}^{*}$ state with $I(J^P)=1/2({3/2}^-)$, $E_2$ is the binding energy of the $\Sigma_c{D}^{*}$ state with $I(J^P)=1/2({1/2}^-)$, and $\Gamma$ is the radiative decay width of the $\Sigma_c D^*|{1}/{2}({1}/{2}^-)\rangle\to \Sigma_c D^*|{1}/{2}({3}/{2}^-)\rangle\gamma$ process with $(I,\,I_3)=(1/2,\,1/2)$.}\label{EEGamma}
\end{figure}

As mentioned above, we adopt different binding energies for the $\Sigma_c{D}^{*}$ state with $I(J^P)=1/2({1/2}^-)$ and the $\Sigma_c{D}^{*}$ state with $I(J^P)=1/2({3/2}^-)$ to discuss the radiative decay width of the $\Sigma_c D^*|{1}/{2}^-\rangle \to \Sigma_c D^*|{3}/{2}^-\rangle \gamma$ process with $(I,\,I_3)=(1/2,\,1/2)$. For the transition magnetic moment of the $\Sigma_c D^*|{1}/{2}({1}/{2}^-)\rangle\to \Sigma_c D^*|{1}/{2}({3}/{2}^-)\rangle\gamma$ process, the obtained numerical results are $1.901\mu_N$ and $1.762\mu_N\sim1.853\mu_N$ in the single channel analysis and the coupled channel analysis, respectively. In Fig. \ref{EEGamma}, we collect the radiative decay width of the $\Sigma_c D^*|{1}/{2}({1}/{2}^-)\rangle\to \Sigma_c D^*|{1}/{2}({3}/{2}^-)\rangle\gamma$ process with $(I,\,I_3)=(1/2,\,1/2)$, which is dependent on the binding energies for the $\Sigma_c{D}^{*}$ state with $I(J^P)=1/2({1/2}^-)$ and the $\Sigma_c{D}^{*}$ state with $I(J^P)=1/2({3/2}^-)$. Here, the transition magnetic moment of the $\Sigma_c D^*|{1}/{2}({1}/{2}^-)\rangle\to \Sigma_c D^*|{1}/{2}({3}/{2}^-)\rangle\gamma$ process with $(I,\,I_3)=(1/2,\,1/2)$ is taken around $1.901\mu_N$, and the binding energies for the $\Sigma_c{D}^{*}$ state with $I(J^P)=1/2({1/2}^-)$ and the $\Sigma_c{D}^{*}$ state with $I(J^P)=1/2({3/2}^-)$ are fixed in the range from $-1~{\rm MeV}$ to $-13~{\rm MeV}$. As shown in Fig. \ref{EEGamma}, the radiative decay width of the $\Sigma_c D^*|{1}/{2}({1}/{2}^-)\rangle\to \Sigma_c D^*|{1}/{2}({3}/{2}^-)\rangle\gamma$ process with $(I,\,I_3)=(1/2,\,1/2)$ is less than $0.06~{\rm keV}$ even if we take the largest transition magnetic moment of the $\Sigma_c D^*|{1}/{2}({1}/{2}^-)\rangle\to \Sigma_c D^*|{1}/{2}({3}/{2}^-)\rangle\gamma$ process with $(I,\,I_3)=(1/2,\,1/2)$, which is similar to the $P_c(4457)^+ \to P_c(4400)^+ \gamma$ decay behavior \cite{Li:2021ryu}. Furthermore, the radiative decay width of the $\Sigma_c D^*|{1}/{2}({1}/{2}^-)\rangle\to \Sigma_c D^*|{1}/{2}({3}/{2}^-)\rangle\gamma$ process with $(I,\,I_3)=(1/2,\,-1/2)$ is less than $0.02~{\rm keV}$. This situation is similar to that for the $P_c(4457)^0 \to P_c(4400)^0 \gamma$ process \cite{Li:2021ryu}.

Before closing this section, we should briefly discuss the relations between the magnetic moments and the transition magnetic moments of the $S$-wave $\Sigma_c^{(*)}D^{(*)}$-type doubly charmed molecular pentaquark candidates. As mentioned in the introduction, these $S$-wave $\Sigma_c^{(*)}D^{(*)}$-type doubly charmed molecular pentaquark candidates may have very short lifetime, which results in the difficulty of  measuring their magnetic moments experimentally. However, we can obtain the transition magnetic moments between the $S$-wave $\Sigma_c^{(*)}D^{(*)}$-type doubly charmed molecular pentaquark candidates extracted by their radiative decay widths, which can be regarded as an indirect way to get the magnetic moments of these possible $S$-wave $\Sigma_c^{(*)}D^{(*)}$-type doubly charmed molecular pentaquarks. Here, the corresponding signs of the magnetic moments and the transition magnetic moments for the $S$-wave $\Sigma_c^{(*)}D^{(*)}$-type doubly charmed molecular pentaquark candidates are  referred to our obtained numerical results.

The magnetic moments of the $S$-wave $\Sigma_c^{(*)}D^{(*)}$ molecular states can be expressed as the linear combination of the transition magnetic moments between the $S$-wave $\Sigma_c^{(*)}D^{(*)}$-type doubly charmed molecular pentaquark candidates, where such combinations are abunandt. In the following, we present an example to discuss the relations between the magnetic moments and the transition magnetic moments of the $S$-wave $\Sigma_c^{(*)}D^{(*)}$ molecules. By simple calculation, we obtain following relation between the magnetic moments and the transition magnetic moments of the $S$-wave $\Sigma_c^{(*)}D^{(*)}$-type doubly charmed molecular pentaquarks with $I_3=1/2$, i.e.,
\begin{equation}
  \left(
    \begin{array}{c}
    \mu_{\Sigma_c D|\frac{1}{2}(\frac{1}{2}^-)\rangle} \\
    \mu_{\Sigma_c^* D|\frac{1}{2}(\frac{3}{2}^-)\rangle} \\
    \mu_{\Sigma_c D^*|\frac{1}{2}(\frac{3}{2}^-)\rangle} \\
    \mu_{\Sigma_c D^*|\frac{1}{2}(\frac{1}{2}^-)\rangle} \\
    \mu_{\Sigma_c D|\frac{3}{2}(\frac{1}{2}^-)\rangle} \\
    \mu_{\Sigma_c D^*|\frac{3}{2}(\frac{1}{2}^-)\rangle}
    \end{array}
  \right) = \left(
    \begin{array}{ccc}
      \frac{5\sqrt{2}}{8} & \frac{3\sqrt{3}}{4} & -\frac{\sqrt{3}}{4} \\
      \frac{3\sqrt{2}}{8} & \frac{9\sqrt{3}}{4} & -\frac{3\sqrt{3}}{4} \\
      -\frac{\sqrt{2}}{8} & \frac{5\sqrt{3}}{4} & -\frac{\sqrt{3}}{4} \\
      -\frac{17\sqrt{2}}{24} & \frac{\sqrt{3}}{12} & -\frac{\sqrt{3}}{4} \\
      \frac{5\sqrt{2}}{8} & \frac{\sqrt{3}}{12} & \frac{5\sqrt{3}}{12} \\
      -\frac{17\sqrt{2}}{24} & \frac{35\sqrt{3}}{36} & -\frac{41\sqrt{3}}{36} \\
    \end{array}
  \right)
  \left(
    \begin{array}{c}
      \mu_1 \\
      \mu_2 \\
      \mu_3
    \end{array}
  \right), \label{relation}
\end{equation}
where we define $\mu_1=\mu_{\Sigma_c^* D|{1}/{2}({3}/{2}^-)\rangle\to \Sigma_c D|{1}/{2}({1}/{2}^-)\rangle}$, $\mu_2=\mu_{\Sigma_c D^*|{1}/{2}({1}/{2}^-)\rangle\to \Sigma_c D|{1}/{2}({1}/{2}^-)\rangle}$, and $\mu_3=\mu_{\Sigma_c D^*|{3}/{2}({1}/{2}^-)\rangle\to \Sigma_c D|{3}/{2}({1}/{2}^-)\rangle}$. Therefore, the magnetic moments of the $\Sigma_c D$ state with $I(J^P)={1}/{2}({1}/{2}^-)$, the $\Sigma_c^* D$ state with $I(J^P)={1}/{2}({3}/{2}^-)$, the $\Sigma_c D^*$ state with $I(J^P)={1}/{2}({3}/{2}^-)$, the $\Sigma_c D^*$ state with $I(J^P)={1}/{2}({1}/{2}^-)$, the $\Sigma_c D$ state with $I(J^P)={3}/{2}({1}/{2}^-)$, and the $\Sigma_c D^*$ state with $I(J^P)={3}/{2}({1}/{2}^-)$ can be accessible by measuring the radiative decay widths of the $\Sigma_c^* D|{1}/{2}({3}/{2}^-)\rangle\to \Sigma_c D|{1}/{2}({1}/{2}^-)\rangle\gamma$, $\Sigma_c D^*|{1}/{2}({1}/{2}^-)\rangle\to \Sigma_c D|{1}/{2}({1}/{2}^-)\rangle\gamma$, and $\Sigma_c D^*|{3}/{2}({1}/{2}^-)\rangle\to \Sigma_c D|{3}/{2}({1}/{2}^-)\rangle\gamma$ processes.

\section{Summary}\label{sec4}

Studying the exotic hadronic states is an intriguing and important research frontier full of opportunities and challenges in hadron physics. After the observation of the charmoniumlike state $X(3872)$, a series of new hadronic states have been reported in different experiments. Explanation of these  observed exotic hadronic states as the hadronic molecules were extensively proposed and studied in the past decades \cite{Brambilla:2019esw,Chen:2016qju,Liu:2013waa,Hosaka:2016pey,Liu:2019zoy,Olsen:2017bmm,Guo:2017jvc,Chen:2022asf,Meng:2022ozq}. In Ref. \cite{Chen:2021kad}, the authors performed the study of the mass spectrum of the $S$-wave $\Sigma_c^{(*)}D^{(*)}$-type doubly charmed molecular pentaquark states. In order to further reveal the properties of these doubly charmed molecular pentaquark candidates, we investigate  their magnetic moments, transition magnetic moments, and radiative decay widths in this work. In our concrete calculation, both the $S$-$D$ wave mixing effect and the coupled channel effect are taken into account.

In the present work, we first focus on the magnetic moments for the $S$-wave $\Sigma_c^{(*)}D^{(*)}$-type doubly charmed molecular pentaquark candidates. By carrying out a quantitative calculation, we conclude: (1) there exist several similarities between the magnetic moments for the $S$-wave $\Sigma_cD^{(*)}$ molecules and that of the $S$-wave $\Sigma_c\bar D^{(*)}$ molecules with the same quantum numbers, (2) the $S$-wave $\Sigma_cD^{(*)}$-type doubly charmed molecular pentaquarks and the compact doubly charmed pentaquarks with the same quantum numbers have different magnetic moments, by which different configurational exotic hadronic states can be distinguished, and (3) the magnetic moment sign of the $\Sigma_cD^{*}$ state with $I(J^P)=1/2(1/2^-)$ is the opposite to that of the $\Sigma_cD^{*}$ state with $I(J^P)=1/2(3/2^-)$, which can be applied to distinguish the spin-parity quantum numbers of the $\Sigma_cD^{*}$ system.
After that, we extend our theoretical framework to study the transition magnetic moments and the corresponding radiative decay widths between the $S$-wave $\Sigma_c^{(*)}D^{(*)}$-type doubly charmed molecular pentaquark candidates. Our numerical results show that the transition magnetic moments between the $S$-wave $\Sigma_c^{(*)}D^{(*)}$-type doubly charmed molecular pentaquark candidates exist much differences, which depend on the flavor-spin wave functions of the initial and final states and the magnetic moments of their components. Moreover, the radiative decay widths between the $S$-wave $\Sigma_c^{(*)}D^{(*)}$-type doubly charmed molecular pentaquark candidates vary from $0$ to $37.061~{\rm keV}$, which strongly rely on the kinetic phase space and the transition magnetic moment.

As a byproduct, we briefly discuss the relations between the magnetic moments and the transition magnetic moments for the $S$-wave $\Sigma_c^{(*)}D^{(*)}$-type doubly charmed molecular pentaquark candidates, which can be regarded as an indirect way to extract the magnetic moments of the $S$-wave $\Sigma_c^{(*)}D^{(*)}$-type doubly charmed molecular pentaquarks in the future experiments.

The magnetic moments and the transition magnetic moments of  the hadronic molecular states play an important role in mapping out their properties. We hope that the present investigation may inspire our colleagues to focus on the electromagnetic properties of the hadronic molecular states in the future. By these efforts, our knowledge of the properties for the hadronic molecular states will become more abundant.

\section*{ACKNOWLEDGMENTS}

H.H.Z. would like to thank Ming-Wei Li for his helpful discussions. This work is supported by the China National Funds for Distinguished Young Scientists under Grant No. 11825503, National Key Research and Development Program of China under Contract No. 2020YFA0406400, the 111 Project under Grant No. B20063, and the National Natural Science Foundation of China under Grant No. 12047501.


\begin{thebibliography}{99}

    \bibitem{Brambilla:2019esw}
    N.~Brambilla, S.~Eidelman, C.~Hanhart, A.~Nefediev, C.~P.~Shen, C.~E.~Thomas, A.~Vairo and C.~Z.~Yuan,
    The $XYZ$ states: Experimental and theoretical status and perspectives,
    \href{https://www.sciencedirect.com/science/article/pii/S0370157320301915?via\%3Dihub}{Phys. Rep. \textbf{873}, 1 (2020)}.

    \bibitem{Chen:2016qju}
      H.~X.~Chen, W.~Chen, X.~Liu, and S.~L.~Zhu,
      The hidden-charm pentaquark and tetraquark states,
      \href{http://linkinghub.elsevier.com/retrieve/pii/S037015731630103X}{Phys.\ Rep.\  {\bf 639}, 1 (2016)}.

    \bibitem{Liu:2013waa}
      X.~Liu,
      An overview of $XYZ$ new particles,
      \href{http://dx.doi.org/10.1007/s11434-014-0407-2}{Chin.\ Sci.\ Bull.\  {\bf 59}, 3815 (2014)}.

    \bibitem{Hosaka:2016pey}
      A.~Hosaka, T.~Iijima, K.~Miyabayashi, Y.~Sakai, and S.~Yasui,
      Exotic hadrons with heavy flavors: $X$, $Y$, $Z$, and related states,
      \href{http://dx.doi.org/10.1093/ptep/ptw045}{Prog. Theor. Exp. Phys. {\bf 2016}, 062C01 (2016)}.

    \bibitem{Liu:2019zoy}
      Y.~R.~Liu, H.~X.~Chen, W.~Chen, X.~Liu, and S.~L.~Zhu,
      Pentaquark and tetraquark states,
      \href{https://www.sciencedirect.com/science/article/pii/S0146641019300304?via\%3Dihub}{Prog.\ Part.\ Nucl.\ Phys.\  {\bf 107}, 237 (2019)}.

    \bibitem{Olsen:2017bmm}
      S.~L.~Olsen, T.~Skwarnicki, and D.~Zieminska,
      Nonstandard heavy mesons and baryons: Experimental evidence,
      \href{https://journals.aps.org/rmp/abstract/10.1103/RevModPhys.90.015003}{Rev.\ Mod.\ Phys.\  {\bf 90}, 015003 (2018)}.

    \bibitem{Guo:2017jvc}
      F.~K.~Guo, C.~Hanhart, U.~G.~Mei$\ss$ner, Q.~Wang, Q.~Zhao, and B.~S.~Zou,
      Hadronic molecules,
      \href{https://journals.aps.org/rmp/abstract/10.1103/RevModPhys.90.015004}{Rev.\ Mod.\ Phys.\  {\bf 90}, 015004 (2018)}.

    \bibitem{Chen:2022asf}
    H.~X.~Chen, W.~Chen, X.~Liu, Y.~R.~Liu and S.~L.~Zhu,
    An updated review of the new hadron states,
    \href{https://arxiv.org/pdf/2204.02649}{arXiv:2204.02649}.

    \bibitem{Meng:2022ozq}
    L.~Meng, B.~Wang, G.~J.~Wang and S.~L.~Zhu,
    Chiral perturbation theory for heavy hadrons and chiral effective field theory for heavy hadronic molecules,
    \href{https://arxiv.org/pdf/2204.08716}{arXiv:2204.08716}.

    \bibitem{Aaij:2019vzc}
      R.~Aaij {\it et al.} [LHCb Collaboration],
      Observation of a narrow pentaquark state, $P_c(4312)^+$, and of two-peak structure of the $P_c(4450)^+$,
      \href{https://doi.org/10.1103/PhysRevLett.122.222001}{Phys.\ Rev.\ Lett.\  {\bf 122}, 222001 (2019)}.

    \bibitem{Li:2014gra}
      X.~Q.~Li and X.~Liu,
      A possible global group structure for exotic states,
      \href{https://link.springer.com/article/10.1140\%2Fepjc\%2Fs10052-014-3198-3}{Eur.\ Phys.\ J.\ C {\bf 74}, 3198 (2014)}.

    \bibitem{Wu:2010jy}
      J.~J.~Wu, R.~Molina, E.~Oset and B.~S.~Zou,
      Prediction of narrow $N^*$ and $\Lambda^*$ resonances with hidden charm above 4 GeV,
      \href{https://journals.aps.org/prl/abstract/10.1103/PhysRevLett.105.232001}{Phys.\ Rev.\ Lett.\  {\bf 105}, 232001 (2010)}.

    \bibitem{Karliner:2015ina}
      M.~Karliner and J.~L.~Rosner,
      New Exotic Meson and Baryon Resonances from Doubly-Heavy Hadronic Molecules,
      \href{https://journals.aps.org/prl/abstract/10.1103/PhysRevLett.115.122001}{Phys.\ Rev.\ Lett.\  {\bf 115}, 122001 (2015)}.

    \bibitem{Wang:2011rga}
      W.~L.~Wang, F.~Huang, Z.~Y.~Zhang and B.~S.~Zou,
      $\Sigma_c \bar{D}$ and $\Lambda_c \bar{D}$ states in a chiral quark model,
      \href{https://journals.aps.org/prc/abstract/10.1103/PhysRevC.84.015203}{Phys.\ Rev.\ C {\bf 84}, 015203 (2011)}.

    \bibitem{Yang:2011wz}
      Z.~C.~Yang, Z.~F.~Sun, J.~He, X.~Liu and S.~L.~Zhu,
      The possible hidden-charm molecular baryons composed of anti-charmed meson and charmed baryon,
      \href{https://iopscience.iop.org/article/10.1088/1674-1137/36/1/002}{Chin.\ Phys.\ C {\bf 36}, 6 (2012)}.

    \bibitem{Wu:2012md}
      J.~J.~Wu, T.~S.~H.~Lee and B.~S.~Zou,
      Nucleon Resonances with Hidden Charm in Coupled-Channel Models,
      \href{https://journals.aps.org/prc/abstract/10.1103/PhysRevC.85.044002}{Phys.\ Rev.\ C {\bf 85}, 044002 (2012)}.

    \bibitem{Chen:2015loa}
      R.~Chen, X.~Liu, X.~Q.~Li and S.~L.~Zhu,
      Identifying exotic hidden-charm pentaquarks,
      \href{https://journals.aps.org/prl/abstract/10.1103/PhysRevLett.115.132002}{Phys.\ Rev.\ Lett.\  {\bf 115}, 132002 (2015)}.

    \bibitem{Chen:2021kad}
    R.~Chen, N.~Li, Z.~F.~Sun, X.~Liu and S.~L.~Zhu,
    Doubly charmed molecular pentaquarks,
    \href{https://doi.org/10.1016/j.physletb.2021.136693}{Phys. Lett. B \textbf{822}, 136693 (2021)}.

    \bibitem{Schlumpf:1993rm}
    F.~Schlumpf,
    Magnetic moments of the baryon decuplet in a relativistic quark model,
    \href{https://journals.aps.org/prd/abstract/10.1103/PhysRevD.48.4478}{Phys. Rev. D \textbf{48}, 4478-4480 (1993)}.

    \bibitem{Ramalho:2009gk}
    G.~Ramalho, K.~Tsushima and F.~Gross,
    A Relativistic quark model for the Omega-electromagnetic form factors,
    \href{https://journals.aps.org/prd/abstract/10.1103/PhysRevD.80.033004}{Phys. Rev. D \textbf{80}, 033004 (2009)}.

    \bibitem{Liu:2003ab}
    Y.~R.~Liu, P.~Z.~Huang, W.~Z.~Deng, X.~L.~Chen and S.~L.~Zhu,
    Pentaquark magnetic moments in different models,
    \href{https://journals.aps.org/prc/abstract/10.1103/PhysRevC.69.035205}{Phys. Rev. C \textbf{69}, 035205 (2004)}.

    \bibitem{Wang:2016dzu}
    G.~J.~Wang, R.~Chen, L.~Ma, X.~Liu and S.~L.~Zhu,
    Magnetic moments of the hidden-charm pentaquark states,
    \href{https://journals.aps.org/prd/abstract/10.1103/PhysRevD.94.094018}{Phys. Rev. D \textbf{94}, no.9, 094018 (2016)}.

    \bibitem{Gao:2021hmv}
    F.~Gao and H.~S.~Li,
    Magnetic moments of the hidden-charm strange pentaquark states,
    \href{https://arxiv.org/pdf/2112.01823}{arXiv:2112.01823}.

    \bibitem{Li:2021ryu}
    M.~W.~Li, Z.~W.~Liu, Z.~F.~Sun and R.~Chen,
    Magnetic moments and transition magnetic moments of $P_c$ and $P_{cs}$ states,
    \href{https://journals.aps.org/prd/abstract/10.1103/PhysRevD.104.054016}{Phys. Rev. D \textbf{104}, no.5, 054016 (2021)}.

\bibitem{Kumar:2005ei}
S.~Kumar, R.~Dhir and R.~C.~Verma,
Magnetic moments of charm baryons using effective mass and screened charge of quarks,
\href{https://iopscience.iop.org/article/10.1088/0954-3899/31/2/006}{J. Phys. G \textbf{31}, no.2, 141-147 (2005)}.

\bibitem{Workman:2022ynf}
R.~L.~Workman \textit{et al.} [Particle Data Group],
Review of Particle Physics,
\href{https://pdg.lbl.gov/#}{PTEP \textbf{2022} (2022), 083C01}.

\bibitem{Huang:2004tn}
P.~Z.~Huang, Y.~R.~Liu, W.~Z.~Deng, X.~L.~Chen and S.~L.~Zhu,
Heavy pentaquarks,
\href{https://journals.aps.org/prd/abstract/10.1103/PhysRevD.70.034003}{Phys. Rev. D \textbf{70}, 034003 (2004)}.

\bibitem{Simonis:2016pnh}
V.~\v{S}imonis,
Magnetic properties of ground-state mesons,
\href{https://link.springer.com/article/10.1140/epja/i2016-16090-5}{Eur. Phys. J. A \textbf{52}, no.4, 90 (2016)}.

\bibitem{Ghalenovi:2021jex}
Z.~Ghalenovi and M.~M.~Sorkhi,
Mass spectra and transition magnetic moments of low lying charmed baryons in a quark model,
\href{https://www.worldscientific.com/doi/abs/10.1142/S0218301321500634}{Int. J. Mod. Phys. E \textbf{30}, no.08, 2150063 (2021)}.

\bibitem{Zhang:2021mam}
W.~X.~Zhang, H.~Xu and D.~Jia,
Masses and magnetic moments of hadrons with one and two open heavy quarks: Heavy baryons and tetraquarks,
\href{https://journals.aps.org/prd/abstract/10.1103/PhysRevD.104.114011}{Phys. Rev. D \textbf{104}, no.11, 114011 (2021)}.

\bibitem{Simonis:2018rld}
V.~Simonis,
Improved predictions for magnetic moments and M1 decay widths of heavy hadrons,
\href{https://arxiv.org/pdf/1803.01809}{arXiv:1803.01809}.

\bibitem{Franklin:1981rc}
J.~Franklin, D.~B.~Lichtenberg, W.~Namgung and D.~Carydas,
Wave Function Mixing of Flavor Degenerate Baryons,
\href{https://journals.aps.org/prd/abstract/10.1103/PhysRevD.24.2910}{Phys. Rev. D \textbf{24}, 2910 (1981)}.

\bibitem{Ozdem:2022vip}
U.~\"Ozdem,
Electromagnetic properties of doubly-heavy pentaquark states,
\href{https://arxiv.org/pdf/2201.00979.pdf}{arXiv:2201.00979}.

\bibitem{Wang:2019spc}
G.~J.~Wang, L.~Y.~Xiao, R.~Chen, X.~H.~Liu, X.~Liu and S.~L.~Zhu,
Probing hidden-charm decay properties of $P_c$ states in a molecular scenario,
\href{https://journals.aps.org/prd/abstract/10.1103/PhysRevD.102.036012}{Phys. Rev. D \textbf{102}, no.3, 036012 (2020)}.

\bibitem{Chen:2017xat}
R.~Chen, A.~Hosaka and X.~Liu,
Searching for possible $\Omega_c$-like molecular states from meson-baryon interaction,
\href{https://journals.aps.org/prd/abstract/10.1103/PhysRevD.97.036016}{Phys. Rev. D \textbf{97}, no.3, 036016 (2018)}.

\bibitem{Chen:2021tip}
R.~Chen,
Strong decays of the newly $P_{cs}(4459)$ as a strange hidden-charm $\Xi _c{\bar{D}}^*$ molecule,
\href{https://link.springer.com/article/10.1140/epjc/s10052-021-08904-4}{Eur. Phys. J. C \textbf{81}, no.2, 122 (2021)}.

\end{thebibliography}
\end{document}